\documentclass[twocolumn,prb,aps]{revtex4-2}

\usepackage{graphicx,color,hyperref}

\usepackage{amsmath}
\usepackage{amsfonts}
\usepackage{amssymb}
\usepackage{dsfont}
\usepackage{braket}
\usepackage{color}
\usepackage{MnSymbol}

\newcommand{\tr}{\mathrm{tr}}

\begin{document}

\title{
Path integral approach to quantum thermalization 
}

\author{Alexander Altland}
\affiliation{Institut f\"ur Theoretische Physik, Universit\"at zu K\"oln, Z\"ulpicher Str. 77, 50937 Cologne, Germany
}

\author{Kun Woo Kim}
\affiliation{Department of Physics, Chung-Ang University, 06974 Seoul, Republic of Korea}

\author{ 
Tobias Micklitz
}
\affiliation{
Centro Brasileiro de Pesquisas F\'isicas, Rua Xavier Sigaud 150, 22290-180, Rio de Janeiro, Brazil 
} 

\date{\today}


\begin{abstract}
We introduce a quasiclassical Green function  approach  describing the unitary yet irreversible
dynamics of quantum systems effectively acting as their own
environment. Combining a variety of  concepts of quantum
many-body theory, notably 
the nonlinear $\sigma$-model of disordered systems, the $G \Sigma$-formalism
for strong correlations, and real time path integration,  the theory is capable of describing a wide range of
 system classes and disorder models. It extends previous work beyond
perturbation theory (in inverse Hilbert
space dimensions), enabling a description of thermalization dynamics  from short
scattering times, through the onset of ergodicity at an effective `Thouless
time', up to
the  
 many-body Heisenberg time. We illustrate the approach with two 
case studies, (i) a brickwork model of unitarily coupled quantum circuits with and without
conserved symmetries, and  (ii)  an array of capacitively coupled quantum
dots. Using the
spectral form factor as a test observable,  we
find good agreement with 
numerical simulations. We present our formalism in a self-contained and
pedagogical manner, aiming  to provide a transferable  toolbox for the  first-principles
description of many-body chaotic quantum systems in regimes of strong entanglement.
\end{abstract}

\maketitle

\section{Introduction}

Describing the irreversible evolution of quantum systems large enough to define
their own  environment has been a long-standing challenge of statistical quantum
physics. The beginning availability of engineered quantum  devices effectively
realizing this setting has led to a surge of renewed efforts to understand the
quantum thermalization problem (see
Refs.~\cite{moriThermalizationPrethermalizationIsolated2018,
francaMakingQuantumDynamics2021,nandkishoreManyBodyLocalizationThermalization2015}
for recent reviews). A crucial aspect of this type of dynamics is that it
manifests itself for  system sizes often beyond the scope of exact
diagonalization, motivating efforts in the development of analytical approaches.
Pioneering contributions in this
direction~\cite{PhysRevX.8.041019,chanSpectralStatisticsSpatially2018,SpectralStatisticsManyBodyConservedCharge2019,fritzschEigenstateThermalizationDualunitary2021}
semiclassically analyze coupled unitary circuits, employing
concepts of statistical mechanics, random matrix theory, and tensor network
theory. 

However, at this point, a first-principles framework, describing 
thermalization in terms of an effective field theory for a wide range of
microscopically different systems and disorder models beyond perturbation theory appears to be missing.
In this paper, we introduce such an approach and illustrate its application on
two different system classes. To establish contact with previous work, the
first  will be realizations of the  standard unitary brickwork design (cf.
Fig.~\ref{fig10}), with and without  continuous symmetries.
Specifically, we will show how structures  beyond perturbation theory in the
inverse of local Hilbert space dimensions, $D$, essentially affect the  dynamics at time scales $t\sim D$.  The
second are arrays of `quantum dots', coupled by energy conserving random
two-body interactions, where energy diffusion turns out to be the essential
bottleneck slowing thermalization.

Modular by design, our approach
starts with the identification of system subunits defined to be 
non-integrable and sufficiently small that they  relax to an ergodic
state  quasi-instantly. The `thermalizing'
systems we are interested in are realized as tensor products of such modules,
coupled by local correlations. 


Conceptually, this our effective description of these systems is a field theory defined over a tensor
product space, as opposed to the more familiar many body field theories in real
space. Its construction draws upon multiple concepts, specifically real time
(Schwinger-Keldysh) path integration~\cite{Altland2023}, quasiclassical Green
functions~\cite{larkin_quasiclassical_1969} the nonlinear $\sigma$-model of
disordered systems~\cite{Efetbook}, the $G
\Sigma$~\cite{rosenhausIntroductionSYKModel2019} (aka
Luttinger-Ward~\cite{luttingerGroundStateEnergyManyFermion1960}) approach to
correlated systems, and insights gained in previous work on thermalization
dynamics~\cite{PhysRevX.8.041019,chanSpectralStatisticsSpatially2018,SpectralStatisticsManyBodyConservedCharge2019,fritzschEigenstateThermalizationDualunitary2021}.
We have tried to present the material in such a way that it is accessible to
readers of different background and interests. The  paper opens with two
non-technical sections, which should  be regarded as an extended
introduction. The first, Section \ref{sec:SummaryOfResults}, discusses  our
qualitative understanding of thermalization dynamics and presents the main
results of this paper in comparison to exact diagonalization. The second,
Section~\ref{sec:Essentials}, introduces conceptual elements of our 
theory construction still in non-technical terms.  In Section \ref{sec:SingleQudit} we introduce our path integral for a
single qudit building block, and in Sections~\ref{sec:CircuitNetworks} and \ref{sec:Symmetries} generalize
it to the brickwork model with and without continuous symmetries, respectively. Energy conserving Hamiltonian
dynamics is discussed in Section 
\ref{sec:HamiltonianSystems}. We conclude in
Section~\ref{sec:Summary}.     

The notation in this paper uses lots of indices, and we aim to be as consistent
as possible. To avoid cluttering, we refrain from using primes, i.e. instead of
$tt'$ we write $tu$, the convention being that second indices are one step up in
the alphabet. For the same reason, we avoid commas, i.e. $\psi_{\mu t}$ instead
of $\psi_{\mu,t}$. For the convenience of the reader, a list of the most
frequently occurring indices is provided in the table below:

\begin{table}[h]
\centering
\resizebox{\columnwidth}{!}{
\begin{tabular}{ll}
\hline
\textbf{Symbol} & \textbf{Meaning} \\
\hline
$a, b$ & Causality indices (retarded/advanced, $+, -$) \\
$c, d$ & Stage indices (e.g., brickwall model stages, $+1, -1$) \\
$\tau,t, u$ & Time indices (discrete or continuous) \\
$\epsilon,\zeta$ & energy variables\\
$m, n$ & Time indices (discrete) \\
$j, k$ & Site (subsystem) indices \\
$\mu, \nu $ & Internal Hilbert space indices (qudit states) \\
$u,v $ & Symmetry representation indices (`spin') \\
$a_j$ & dimensionless time translation parameters\\
$q$ & Momentum (Fourier) mode index \\
$D$ & Local Hilbert space dimension (qudit dimension) \\
$L$ & Number of sites/subsystems \\
$\lambda$ & High energy cutoff\\
$\Gamma $ & Interaction damping rate\\
\hline
\end{tabular}
}
\caption{Summary of the most frequently occurring symbols and index conventions used throughout the paper.}
\label{tab:indices}
\end{table}

\section{Summary of concepts and results}
\label{sec:SummaryOfResults}

We begin our discussion by  reviewing the
concept of 
quasiclassical Green functions as  slowly fluctuating variables
describing the late time stages of chaotic evolution. This will be followed by a
summary of our results for systems with discrete and continuous time evolution,
respectively. 

\begin{figure*}
    \centering
    \includegraphics[width=0.9\linewidth]{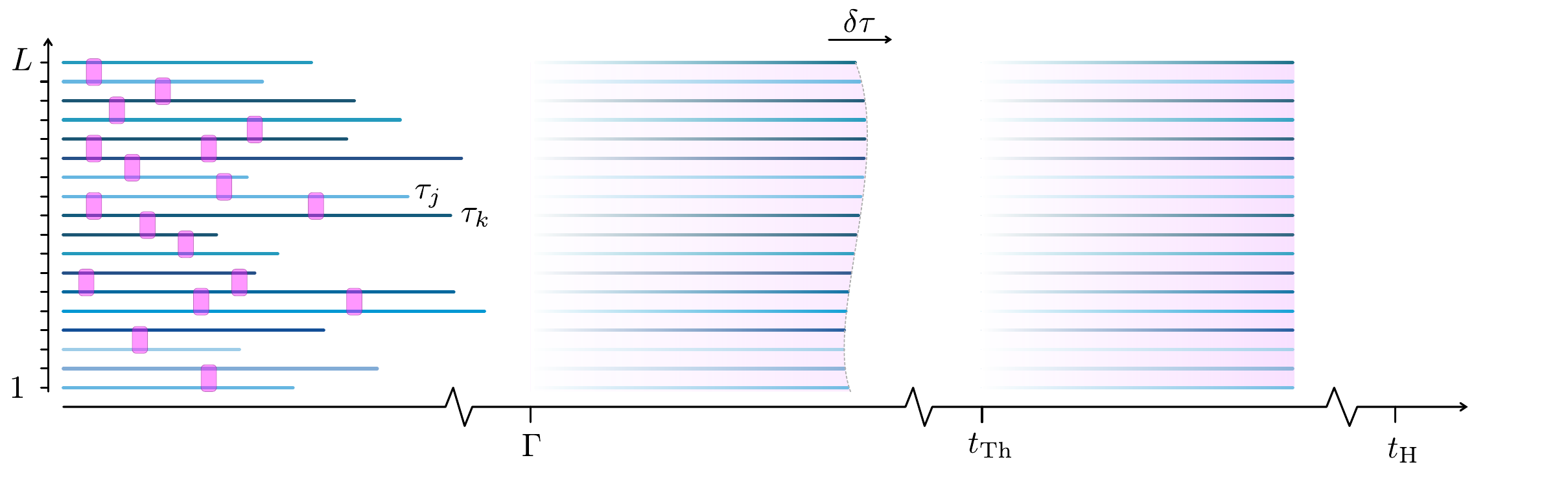}
    \caption{The three stages of quantum thermalization in an energy conserving 
    medium defined by  $L$ chaotic, statistically independent 
    (symbolically indicated by the color scheme) 
    and pairwise entangled subsystems. Left:
    For times shorter than a characteristic pair interaction scale, 
    $\Gamma$, the system supports $L$ dynamical zero modes: 
    the differences in time $\tau_j-\tau_k$ at which subsystem quantum 
    amplitudes propagate in their respective Hilbert spaces. 
    Center: At time scales exceeding $\sim \Gamma \ln(L)$, 
    global phase coherence is maintained only by states with small time differences 
    $\Delta \tau\sim\tau_j-\tau_k$. 
    However, time translational invariance (due to energy conservation) implies
    that weak collective fluctuations in $\delta \tau$, remain a soft mode. Right: Beyond an effective `Thouless time' 
    $\tau_\textrm{Th}\sim DL^2$, these modes have relaxed, too, and the 
    system enters a late time ergodic regime. }
    \label{fig1}
\end{figure*}

\subsection{Quasiclassical Green functions}
 
Consider a $D$-dimensional quantum system which is strongly chaotic in the sense
that an ergodic state is reached after microscopically short time scales.
Examples of such building blocks include
\begin{itemize}
    \item Irregularly shaped or disordered quantum dots,
    \item Systems described in terms of unitarily invariant random matrix
    Hamiltonians, or 
    \item Qu$D$its evolving under Haar distributed unitaries. 
\end{itemize}
We will summarily refer to these units as \emph{qudits} throughout.
Time scales below which  qudits become ergodic are not resolved by our
theory.  

We describe individual qudits  in terms of collective variables 
\begin{align}
    \label{eq:QuasiaClassicalGreenFunction}
       -\frac{ia}{D} \sum_\mu \psi^a_{\mu t}\bar \psi^{b}_{\mu u}\equiv  G^{ab}_{tu},
\end{align}
where $\mu$ runs over a qudit's  $D$ internal states  and  $\psi_{\mu,t}^a$ are time dependent
fields~\footnote{Within our later path integral approach, their role is taken by
Grassmann integration variables, hence the denotation `fields'. Alternatively,
one may think of them as fermionic or bosonic creation and annihilation operators.} of either retarded, $a=+1$,
or advanced, $a=-1$ causality. These variables afford different interpretations:
Considered as building blocks in correlation functions, and before ensemble
averaging, they are defined through their time dependence
\begin{align}
    \label{eq:PairCorrelationFunction}
    \left \langle \psi^a_{\mu t}\bar \psi^{b}_{\nu u} \right \rangle 
    = -i \Theta(a(t-u)) \delta^{ab}\braket{\mu|e^{i  (t-u) (-H+i\delta s) }|\nu}.
\end{align}
(For a system with unitary evolution replace
$e^{-i H t}\to U^t$, with $t=n\in \Bbb{Z}$.)
In this reading, the fields represent
wave functions $\psi^\pm_{\mu,t}\to \ket{\mu(\pm t)}$ propagating forward or
backward in time.  Under the
assumed condition of ergodicity, individual $\psi_\mu$ are rapidly fluctuating
amplitudes. However, the summation over $\mu$, makes the pairings 
$G\sim \psi_\mu \bar{\psi}_\mu$ good candidates for slowly fluctuating effective
variables. 

When inserted as building blocks in correlation functions, the $G_{tu}$  remain slowly fluctuating, including
for macroscopic differences $\tau\equiv t-u$ between the readout times.
To illustrate this principle, consider the spectral form factor (SFF)
\begin{align*}
    K(t)\equiv\left \langle \left| \tr(e^{iHt})\right|^2 \right \rangle_H,
\end{align*}
where the angular brackets on the right-hand side denote an average over an
ensemble of Hamiltonians. With Eq.~\eqref{eq:PairCorrelationFunction}, it is
straightforward to obtain the representation
\begin{align}
    \label{eq:FormFactorSingle}
    K(t)= D^2 \left \langle G^{++}_{t0} G^{--}_{0t} \right \rangle,
\end{align}
where the brackets on the r.h.s. now include an  integral over the
$\psi$-variables contained in $G$\footnote{Depending on the context $\langle \dots \rangle$ will
denote path integration, ensemble averaging, or both. In cases, of ambiguity, we
occasionally denote the latter by $\langle \dots \rangle_H$.}. In this case, the
Green functions are evaluated for time differences equaling the observation time
of the system. We note that  quasiclassical Green functions have been
instrumental in  various contexts, including the Luttinger-Ward
functional~\cite{luttingerGroundStateEnergyManyFermion1960} (aka
$G\Sigma$-functional~\cite{rosenhausIntroductionSYKModel2019}), the nonlinear
$\sigma$-model~\cite{Efetbook},   or  in the context of
disordered superconductivity~\cite{larkin_quasiclassical_1969}.

 For an $L$-qudit system, we  have as many
collective variables, $G_j$, realized in the tensor product structure defined by
its Hilbert space. For each of them, the time differences $t_j-u_j\equiv \tau_j$
are `soft variables', as schematically indicated in
Fig.~\ref{fig1}, left.  Turning on inter-qudit correlations, we
ask which of these configurations
contribute to the path integral in the long time limit.  
Interactions are instantaneous in time, which 
will turn out to constrain the freedom to choose the time differences $\tau_j$
independently. Rather, we will find that correlations confine the mismatch $\Delta
\tau=\tau_j-\tau_k$ between the time variables of neighboring qudits to be
microscopically small, Fig.~\ref{fig1}, center. For time
scales larger than an effective `Thouless time', $t_\textrm{Th}$, even these
differences get damped out and the collective amplitudes supporting the system's
dynamics assume the form
\begin{align}
    \label{eq:GHadamard}
  &  G_{1 t_1 u_1}^{a_1 b_1} \dots G_{L t_L u_L}^{a_L b_L}\to G_{1 t u}^{ab}... G_{Lt u}^{ab}\equiv\cr
    &\qquad\equiv  (G_1 \odot ... \odot G_L)^{ab}_{tu}\equiv G^{ab}_{tu},
\end{align} 
i.e. a freezing to one common time set of time and  causal indices
where the notation on the right-hand side uses the element wise, or Hadamard
product of matrices
\begin{align}
    \label{eq:HadamardProduct}
    (A\odot B)_{tu}^{ab}\equiv A_{tu}^{ab}B_{tu}^{ab}.
\end{align}
Physically, this state reduction reflects a `synchronization' of $L$
independently propagating  states to a single time-dependent
many-body wave function. Their effective quasiclassical Green function $G$ then
describes the late time physics in its ergodic (`thermalized') regime, cf.
Fig.~\ref{fig1}, right.

\begin{figure*}
    \centering
  \includegraphics[width=0.32\textwidth]{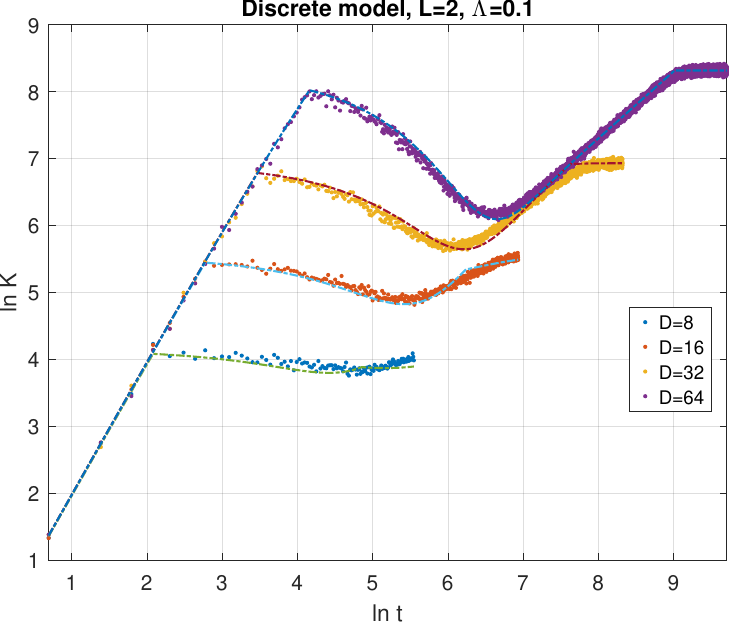} 
  \includegraphics[width=0.32\textwidth]{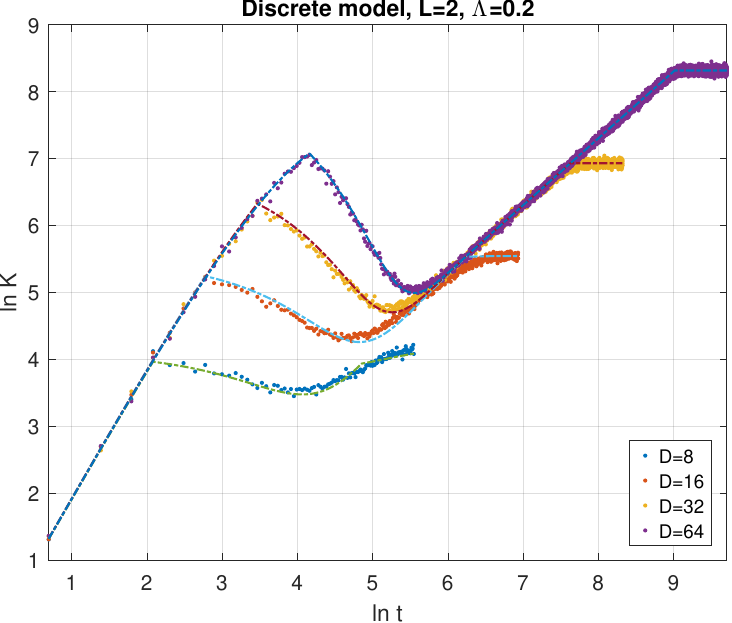} 
  \includegraphics[width=0.32\textwidth]{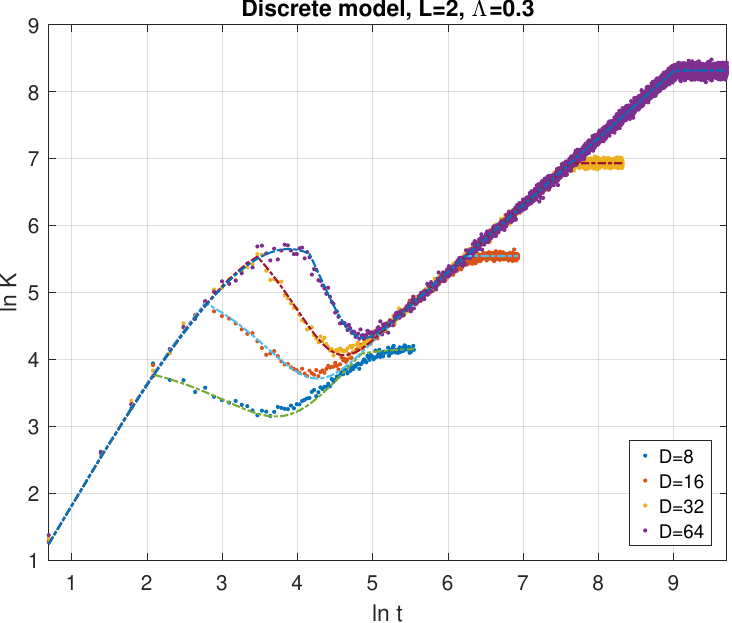} 
    \caption{ The SFF of a two-site system, $L=2$, for three different
  interaction strength, $\Lambda=0.1$ (left) $\Lambda=0.2$ (middle)
  $\Lambda=0.3$ (right). Each data set is averaged over $10^3$ realizations. We
  fit these results with the effective damping rate  $\Gamma$ as a single
  parameter, where, for qudit dimensions $D=8,16,32,64$, 
$\Gamma= 0.0039, 0.0028, 0.0022, 0.0023$  
($\Lambda=0.1$), 
$\Gamma= 0.010,    0.0085,    0.0090,    0.0095$ 
($\Lambda=0.2$) 
$\Gamma= 0.020,    0.020,    0.0210,    0.0215$  ($\Lambda=0.3$).  
     }
    \label{fig2}
\end{figure*}

\subsection{Discrete time evolution}

As a  simple model exemplifying this type of dynamics, we consider the
brickwork design illustrated in 
Fig.~\eqref{fig10}. It contains horizontal rows of $L$
statistically independent qudits which are fast-thermalizing in the sense that
time evolution $\exp(-iH_j \delta)\equiv U_j$ is statistically equivalent to
operators drawn from a Haar distribution. Correlations between neighboring
qudits, $j,k$ are described by interaction operators $\exp(i\delta H_{j,k})$
with Gaussian distributed interaction operators of variable strength,
Eq.~\eqref{eq:InteractionVariance}.  

For this system, we average over realizations of qudit operators to derive a
real-time path integral over quasiclassical Green functions. To leading order in
an expansion in $D\gg 1$ and $t\gg 1$, it yields the SFF as
    \begin{align}
    \label{eq:FormFactorPotts}
    K(t)&\simeq \left( 1 +t e^{-\Gamma t}\right)^L + t\left( 1-e^{-\Gamma t}  \right)^L,
\end{align}
identical to that of a somewhat different brickwork model in
Ref.~\cite{chanSpectralStatisticsSpatially2018}. In this formula, $t$ is
discrete time, and $\Gamma=\frac{\Lambda^2}{4}$  a dimensionless golden-rule
interaction rate determined by the square of the interaction matrix elements,
$\Lambda^2/D^2$, multiplied by the two-qudit spectral density $\sim D^2$.
Eq.~\eqref{eq:FormFactorPotts} describes a crossover from the semiclassical form
factor $K(t)= t^L$ of $L$ independent qudits in the limit of absent interaction
$\Gamma t\to 0$, to that of a single ergodic system $K(t)\simeq t$ for $\Gamma t
\gg 1$. Governed by the  factors $\exp (-\Gamma t)$, this crossover is
exponentially fast, and completed after time scales $\sim \Gamma^{-1}\ln L$.

We also note that Eq.~\eqref{eq:FormFactorPotts} affords a statistical mechanics interpretation as
the partition sum of a $t$-state Potts
model~\cite{chanSpectralStatisticsSpatially2018}. Within it, the interaction of
$t$-state `spins' $s_j\in \Bbb{Z}_t$  flags whether two neighboring discrete
time variable (differences) are in synchronization ($s_j=s_k$) or not. Summation
over these configurations yields Eq.~\eqref{eq:FormFactorPotts}.

However, for weaker interactions $\Gamma t \lesssim 1$ the
semiclassical/statistical mechanics approach
no longer adequately describes the crossover dynamics. We now need to take into
account that the ergodic form factor of $n$ ergodically coupled qudits is not
just linear in $t$, but  given by~\cite{Haake}
\begin{align}
    \label{eq:HaakeFormFactor}
    K_n(t)= t\, \Theta(D^n-t)+ D^n \,\Theta (t-D^n).
\end{align} 
Within the path integral formalism, information about the late time stationary
values (the `plateau regime') is contained in saddle point configurations known
as Altshuler-Andreev (AA) saddles~\cite{andreevSpectralStatisticsRandom1995}. Their
inclusion leads to the generalized result
     \begin{align}
    \label{eq:FormFactorDiscreteNonPert}
    K(t)=\left(1+K_1(t)e^{-\Gamma t}\right)^L+K_L(t)\left(1-e^{-\Gamma t}\right)^L
 \end{align}
While the departure from linearity in $K_L$ kicks in
 only at very late times $t>D^L$, the saturation to a plateau value in $K_1$
 takes place at $t=D$, i.e. time scales easily competing with inverse interaction rates
 $\sim \Gamma^{-1}$. As we are going to discuss next, the inclusion of this structure makes a significance
 difference in the comparison to exact diagonalization.

 \paragraph*{Numerical validation:} Fig.~\ref{fig2} shows the results of simulations for
$L=2$, qudit dimensions $D=8,16,32,64$, and various two-qudit interaction
strength, $\Lambda$. The agreement with the analytical
formula (green) gets better with increasing $D$, when 
$D^{-1}$-corrections beyond our approach  diminish. For the largest values of $D$, the
one fit parameter used in this comparison, the damping rate $\Gamma=\Lambda^2/4$ equals the
analytical prediction, i.e. the comparison with the numerical results is
essentially parameter free. Also notice that the difference between the
approximation Eq.~\eqref{eq:FormFactorPotts} and
Eq.~\eqref{eq:FormFactorDiscreteNonPert} --- visible in the kink-like structures
at the Heisenberg times $D$ of the individual qudits --- is essential in
obtaining quantitative agreement. 

\begin{figure}
    \centering
  \includegraphics[width=0.4\textwidth]{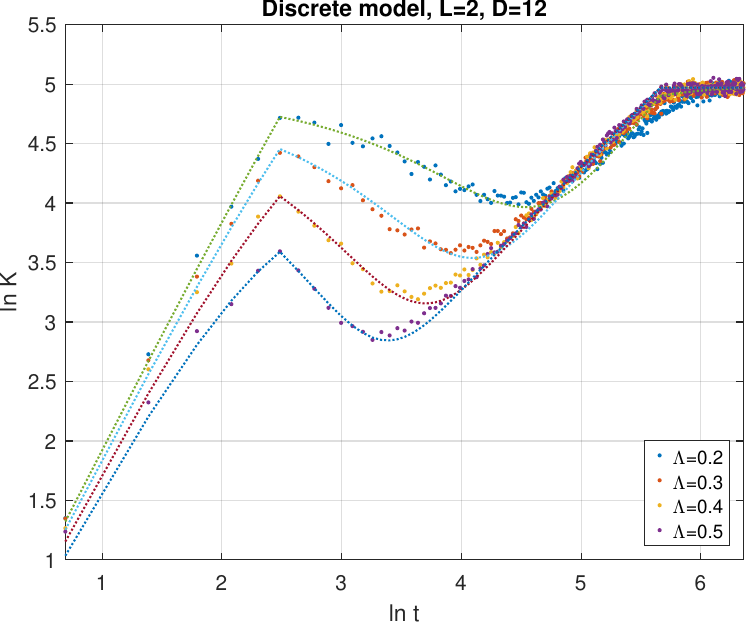} \\
  \includegraphics[width=0.4\textwidth]{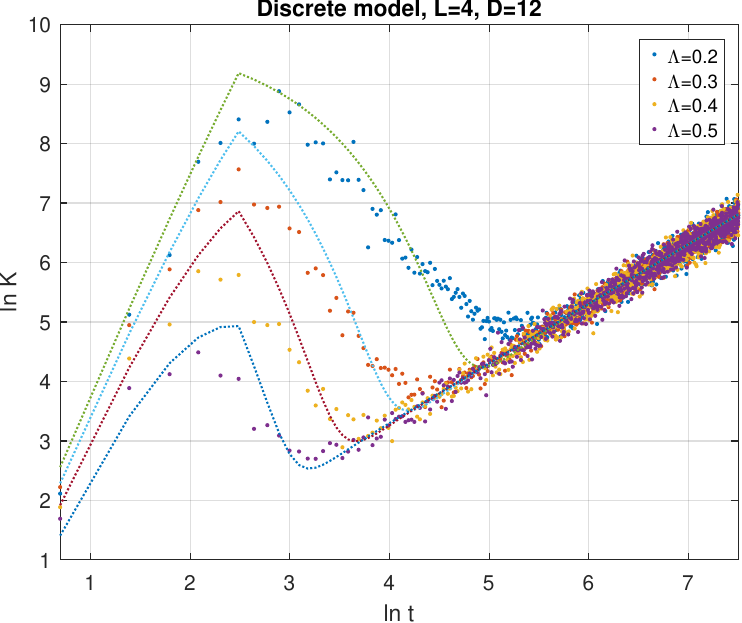} 
    \caption{
  The SFF for $D=12$, interaction strengths, $\Lambda=0.2,0.3,0.4,0.5$ and
  system sizes $L=2$, $10^3$ runs  (top) and $L=4$, $200$ runs  (bottom). For increasing interaction
  strength, we use the fitting parameters 
$\Gamma=0.009,    0.019,    0.034,    0.053$ ($L=2$) and 
$\Gamma=0.0135,    0.0310,    0.0550,    0.0900$, ($L=4$).  
}
    \label{fig3}
\end{figure}

The
largest arrays we have analyzed have $L=4$ with dimension $D=12$, 
Fig.~\ref{fig3}. Comparing to the case $L=2$ with the same dimension,
shown in the left panel for comparison, we observe sizeable deviations off
analytical formula. We also note that for large interactions the best fits are obtained for
 damping rates $\Gamma$, significantly larger than the simple
golden rule estimates by factors of $\mathcal{O}(1-2)$, which may hint at an
effectively larger scattering
phase volume  in the multi-site array. However, considering the smallness of the dimension $D=12$
the agreement with a theory having $D^{-1}$ as an expansion parameter, is still satisfactory. 

\paragraph*{Discrete time evolution with symmetries:}
We mentioned in the introduction that  local
conservation laws reflecting the presence of a symmetry  dramatically slow down the relaxation into ergodic
states\cite{SpectralStatisticsManyBodyConservedCharge2019}. The first stage of
thermalization outlined above then is followed by a
slower second one governed by the (typically diffusive) exchange dynamics of conserved
charges. 

As a simple model
 for such a scenario, we consider the previous one, upgraded to a state space $\mu \to (\mu,a)$ where $a=1,\dots,N$ are the globally
conserved `charges' of the symmetry, think of the $z$-quantum number of a spin,
etc. Symmetry transformations with weak site-to-site variations
describe the relaxation of the symmetry's
charges. As pointed out in
Ref.~\cite{SpectralStatisticsManyBodyConservedCharge2019} in the context of $\mathrm{SU}(2)$-symmetry, the diffusive nature
of this process implies full
equilibration on time scales
$\sim L^2$. In Section~\ref{sec:Symmetries} we include this mechanism into our
path integral in terms of a (non-random) $\mathrm{SU}(N)$-spin exchange operator.

As a result, we obtain a find the form factor at time scales exceeding $\sim
\Gamma \ln(L)$ modulated by a  factor describing the time evolution of an
$\mathrm{SU}(N)$-Heisenberg model (cf. Eg.~\eqref{eq:SFFFormFactorSymmetries}).
The late time dependence of the latter is set by the action cost $\sim t L^{-2}$
of its lowest lying magnon excitations, establishing a time scale $t \sim
L^{2}$. (The fact that  the magnetic action
enters with a purely imaginary coupling is due to the specific modeling of
the spin exchange. A generalization to randomly fluctuating spin operators would
lead to relaxation.)

\subsection{Continuous time evolution}

In Section~\ref{sec:HamiltonianSystems}, we consider linear arrays of
capacitively coupled quantum dots as an example of an energy conserving system.
In this case, time translations $\tau_j \to \tau + \lambda^{-1}a_j$ controlled
by small dimensionless parameters $a_j = \mathcal{O}(1)$  assume the role of
symmetry generators, and Eq.~\eqref{eq:ContinuousTimeAAction} assumes a role
analogous to that of the Heisenberg action, in the symmetry-enriched discrete
time framework.    
One technical issue  is that the deviations in the $a$-paramaters from site to
site are small, of the order of the inverse of the high energy cutoff $\lambda$
of the theory. Our effective low energy theory lacks precision control over
these scales, meaning that the action is obtained with parametric, but not
numerical accuracy.

Integrating over $a$-modes we obtain 
\begin{align}
    \label{eq:ContinuumProductFormula}
    K(t)\sim C t \prod_{n=1}^{L-1} F_n, \qquad F_n = \max\left(\left(t \Gamma \sin^2 \left( \frac{\pi n}{L} \right) \right)^{-1/2},1\right),
\end{align}
in the regime $t\gg \Gamma$, where $C$ is a numerical constant. The qualitative difference to the previously
derived Eq.~\eqref{eq:FormFactorPotts} is the soft relaxation
towards the ergodic limit on time scales $t\sim L^2$.

\begin{figure}
    \centering
    \includegraphics[width=0.4\textwidth]{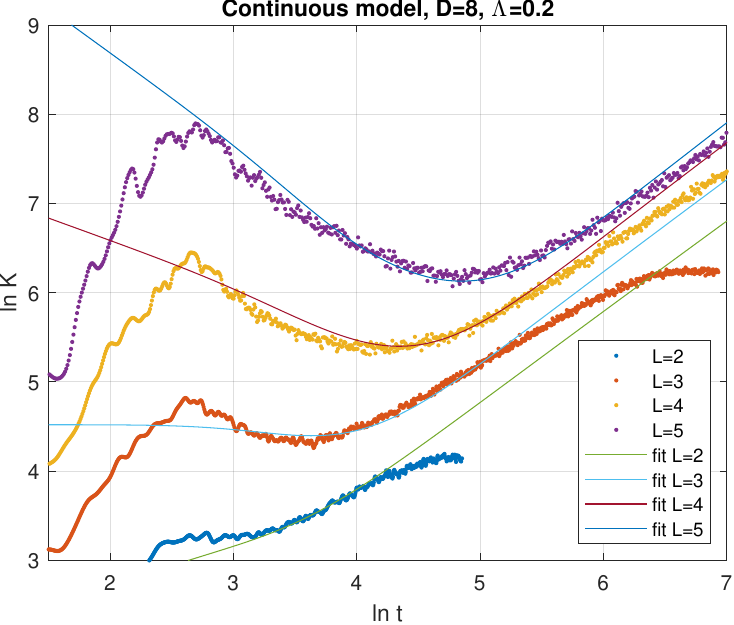}
    \caption{The product formula Eq.~\eqref{eq:ContinuumProductFormula} compared
   to numerics for systems of size $L=2,3,4,5$ with $D=8$. We fit the numerically obtained SFF with
    $\Gamma \equiv \Lambda^2/2 \lambda= 0.024,0.0210,0.0178,0.0152$ 
    for increasing length. 
    A second fit parameter
 $C= 0.86,1.45,2.34,3.16$ is used to fit the undetermined constant of
 proportionality in Eq.~\eqref{eq:ContinuumProductFormula}.
 }
    \label{fig4}
\end{figure}

\paragraph*{Numerical validation:}  Figure \ref{fig4} compares
Eq.~\eqref{eq:ContinuumProductFormula} to numerical simulations of arrays of
length $L=2,3,4,5$ with low Hilbert space dimension $D=8$. Our lack of control
over  high energy scales requires working with two fit parameters. The first is
the overall multiplicative constant $C$, and the second the parameter $\Gamma$.
 We also used a
 $\mathrm{max} $-function softened over $\lambda$-scales.

\section{Path integral essentials}
\label{sec:Essentials}

Having summarized our main findings, we now turn to their derivation. As
mentioned in the introduction, our principal workhorse will be a path integral over quasiclassical Green functions describing
the evolution of correlated states in a tensor product space.  To guide its
construction, we begin by discussing the 
physics of these $G$-variables from two complementary perspectives:
semiclassical analysis, and symmetries:

\subsection{Semiclassical analysis} 
\begin{figure}[h]
    \centering
    \includegraphics[width=1\linewidth]{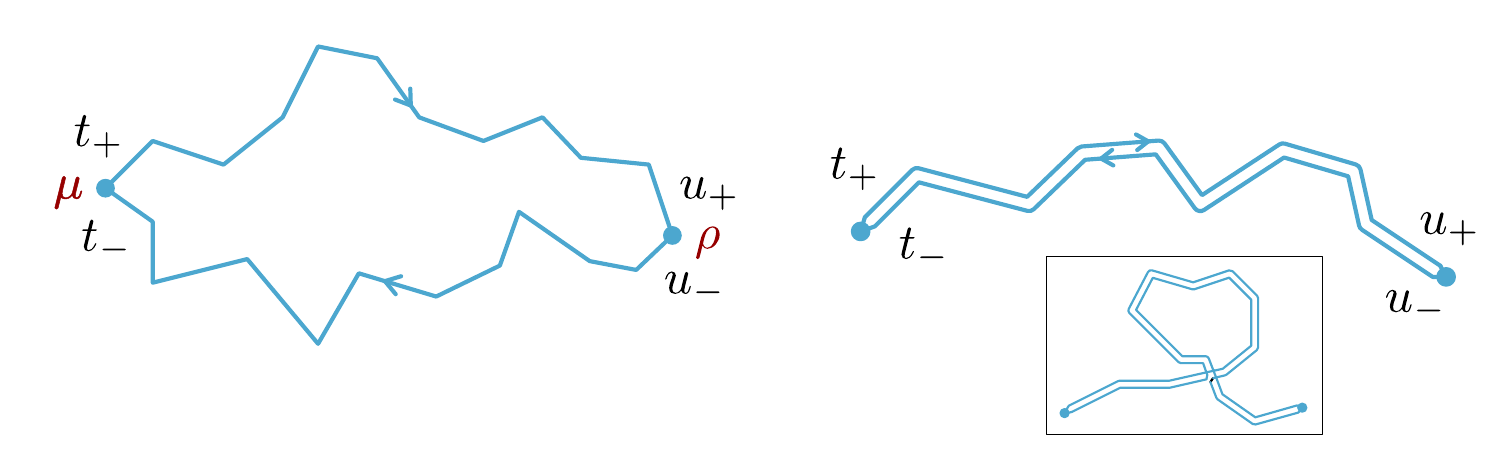}
    \caption{Left: coherent double sum over retarded (top) and advanced (bottom) scattering path propagators connecting two qudit states $\mu$ and $\rho$. Left: Averaging removes the contribution of different paths leading to a semiclassical pair propagator and constraints on the propagation times.}
    \label{fig5}
\end{figure}

Consider the correlation function 
\begin{align*}
    \Pi_{u_+ u_-,t_+ t_-}&\equiv \langle G^{+-}_{u_+u_-}G^{-+}_{t_-t_+}
\rangle =\\ 
& =\frac{1}{D^2} \left  \langle \psi^+_{\rho u_+}\bar \psi^+_{\mu t_+}\  \psi^-_{\mu t_-}\bar \psi^-_{\rho u_-}   \right \rangle
\end{align*}
of two Green functions,
Eq.~\eqref{eq:PairCorrelationFunction}, where  we have reordered field variables to highlight an
interpretation  as the product of a retarded and an advanced transition
amplitude $\ket{\rho} \to \ket{\mu}$ in time $t_+-u_+$ and $t_--u_-$, respectively, Fig.~\ref{fig5}, left.    
Considering  these amplitudes individually expanded as coherent sums over
Feynman scattering paths, the majority of path configurations to the double sum
(such as the one shown on the left) will average out due to random phase
cancellations. To leading order in $D^{-1}$, the dominant contribution to the
product $\Pi$ is provided by pairs of identical paths, whose retarded and
advanced amplitudes are complex conjugate to each other.  The argument is
`semiclassical' in that quantum interference `loop corrections' (inset) will
begin to matter at higher orders in a $D^{-1}$ expansion. They are included in
our construction below, but for the moment we concentrate on the leading
semiclassical order. 

For a system with a well definite classical limit, this pair propagator defines
the semiclassical representation of a classical transition process between
different states $\rho$ and $\mu$. Ergodicity implies its independence of both the
choice of $\mu$ and $\rho$, and of the propagation time $u_+-t_+$. At the same time,
the rigid locking between the advanced and retarded scattering path requires the
equality of propagation times $u_+-t_+=u_--t_-$, or $u_+-u_-=t_+-t_-$. We thus
hypothesize 
\begin{align}
    \label{eq:GpmPropagator}
    \Pi_{u_+ u_-,t_+ t_-}=\left \langle G_{u_+ u_-}^{+-} G_{t_- t_+}^{-+} \right \rangle=\frac{1}{D \lambda} \delta_{\Delta t-\Delta u}\Theta\left(u- t \right).
\end{align}
Here, and throughout, $\delta_t$ are $\delta$-functions smeared over the inverse
of the cutoff scale, $\lambda$, i.e. $\delta_0=\lambda$. For discrete time dynamics,
$\lambda=1$ and $\delta$ is a Kronecker-$\delta$.  
The Heaviside function implements the causality conditions $u_s > t_s$ and
we introduced center of mass coordinates, $\Delta t = t_+-t_-$ and
$t=(t_++t_-)/2$. The normalizing prefactor follows from probability conservation
$ \Pi_{uu,tt}=1$.

\subsection{Symmetries} 
\label{eq:EssentialsSymmetries}

Eq.~\eqref{eq:GpmPropagator} describes the 
evolution of the $G^{\pm \mp}$ components of the quasiclassical Green function.
To obtain complementary information for the diagonal blocks, $G^{\pm
\pm}$, we now turn to symmetry arguments, here formulated in the
continuous time framework for concreteness. (All conclusions carry over to
discrete time evolution without changes.) 

The present approach to quantum chaos understands the
ergodic phase as a symmetry broken phase. The symmetry in question is that
between the retarded and advanced dynamical generators $H\pm i\delta$, which are
identical up to the infinitesimal causality increment $\pm i\delta$. In the
energy domain, Fourier transform of  Eqs.~\eqref{eq:PairCorrelationFunction}
(according to the convention $ f_\epsilon = \int dt \,e^{i\epsilon t} f_t$) yields
\begin{align*}
    \langle G^{ab}_{\epsilon \zeta}\rangle 
    = \delta^{ab} \delta_{\epsilon -\zeta} \frac{1}{D}\sum_\mu \braket{\mu |\frac{1}{\epsilon + ia \delta -H }|\mu},
\end{align*} 
i.e. an expression which, for energies $\epsilon,\zeta$ away from the poles
of $H$ (a discrete set of measure zero embedded into the continuum of energies),
is independent of the sign of $a$. We may express this independence as the statement that
the theory possesses a continuous  symmetry under transformations $G\to TG
T^{-1}$, where $T$ is a two-dimensional matrix acting on the
$a$-indices.  

This symmetry is spontaneously broken by the configuration average in the sense that
\begin{align*}
    \left \langle G^{++}_{\epsilon \zeta} - G^{--}_{\epsilon \zeta}\right \rangle_H&=  \delta_{\epsilon-\zeta} 2i \frac{\textrm{Im}}{D}\sum_\mu  \left \langle \braket{\mu |\frac{1}{\epsilon + ia \delta -H }|\mu} \right \rangle_H\cr 
   & \equiv -\frac{2 \pi i}{D}  \delta_{\epsilon-\zeta}\rho(\epsilon),
\end{align*} 
i.e. a non-vanishing difference between the averaged retarded and advanced Green
function, proportional to the average density of states,
$\rho(\epsilon)=\Delta(\epsilon)^{-1}$, or the inverse of the level spacing. The
average $\langle G \rangle_H$  `spontaneously breaks' the unitary invariance
between the two causal sectors. 

However,  as we are  in an effectively zero-dimensional framework (ergodicity),
we  expect Goldstone mode fluctuations to restore the symmetry via strong
fluctuations. We also expect these fluctuations to become effective in the limit
of times $t \gg  t_\textrm{H}\sim \Delta^{-1}$ larger than the Heisenberg time.
In this long time regime, the fine-grained pole structure of the spectrum on
scales $\Delta$ is resolved, and the symmetry between the retarded and advanced
sector of the theory must become visible again. 

For times $t\gg \lambda^{-1}$, we expect these Goldstone modes to be the
dominant fluctuation degrees of freedom. We include them into the theory by
parameterizing the quasiclassical Green function --- now interpreted as an
effective integration variable ---   as 
\begin{align}
    \label{eq:GvsQ}
    G^{ab}_{tu}\to - \frac{i}{\lambda} (T \tau_3 T^{-1})^{ab}_{tu}\equiv -\frac{i}{\lambda}\,  Q^{ab}_{tu},
\end{align}
(in
the discrete case, $\lambda=1$) where $(\tau_3)_{ab}=a \delta_{ab}$ is a Pauli
matrix in  causal space. Finally, $T=\{T_{tu}^{ab}\}\in \mathrm{U}(2N)$ are
generalizations of the exact two-dimensional symmetry matrices to bi-local
matrices exhibiting slow dependence on time indices, where we consider time
discretized into $N$ steps. The saddle point $\propto \tau_3$ breaks this
symmetry down to $\mathrm{U}(N)\times \mathrm{U}(N) $, and the matrices $Q$ span
the associated Goldstone mode manifold $\mathrm{U}(2N)/ \mathrm{U}(N)\times
\mathrm{U}(N) $. In this way they resemble $2N$-dimensional `spin'
configurations continuously rotated away from the mean field $\tau_3$. 

\begin{figure}[h]
    \centering
    \includegraphics[width=0.9\linewidth]{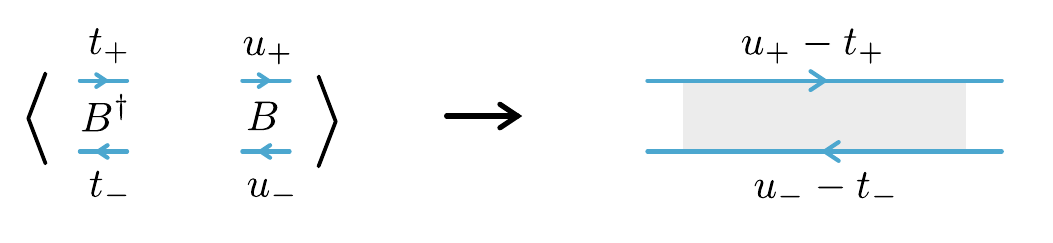}
    \caption{Left: identification of Goldstone mode generators with the paired fields entering the quasiclassical Green function. Right: the pairing of Goldstone mode generators identified with the ergodic pair propagator with propagation times $u_+-t_+=u_--t_-$.}
    \label{fig6}
\end{figure}

\subsection{Low energy fluctuations}

In concrete terms, we represent fluctuations  away from $T=\mathds{1}$ in terms
of the
so-called rational parameterization~\cite{Efetbook}
\begin{align}
    \label{eq:RationalParameterization}
    T=\begin{pmatrix}
    \mathds{1}& B\cr -B^\dagger & \mathds{1}
    \end{pmatrix},
\end{align}  
where $B=\{B_{tu}\}$. The matrix  $T^{-1}$ then assumes the form of a geometric
series, and  a straightforward resummation  leads to 
\begin{align}
    \label{eq:QMatrixB}
    Q=\begin{pmatrix}
        \frac{1-B B^\dagger}{1+B B^\dagger}& - \frac{2B}{1+B^\dagger B}\cr
        - \frac{2B^\dagger}{1+B B^\dagger}&- \frac{1-B^\dagger B}{1+ B^\dagger B}
    \end{pmatrix}.
\end{align} 
(Readers familiar with spin coherent
states~\cite{perelomov2002coherentstatesarbitrarylie} may notice the resemblance
with a spin degree of freedom represented in stereographic projection,
underscoring the magnetic analogy.) 

A truncation at second order in
Goldstone modes yields the Green function as
\begin{align}
    \label{eq:GinBExpansion}
    G_{tu}=-\frac{i}{\lambda} \begin{pmatrix}
    \mathds{1}-2B B^\dagger & -2B\cr 
    -2B^\dagger &-\mathds{1}+2B^\dagger B
    \end{pmatrix}_{tu}+\dots,
\end{align}
an expression  establishing contact to the previously discussed semiclassical picture:

\begin{figure}[h]
    \centering
    \includegraphics[width=0.6\linewidth]{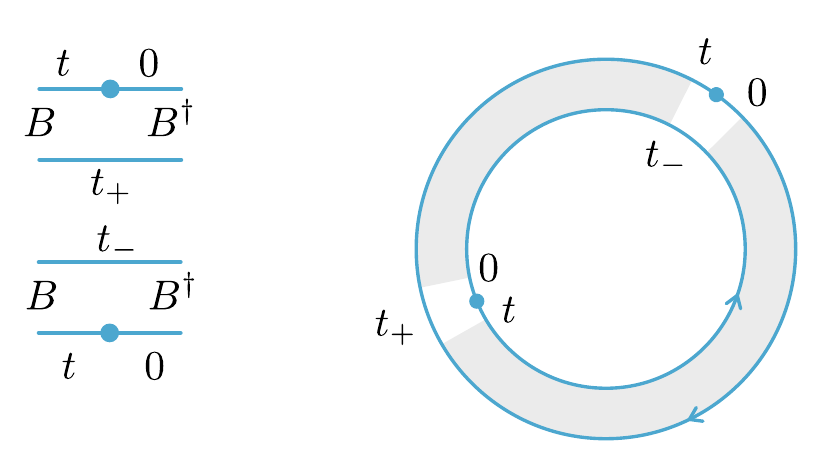}
    \caption{The pairing of four Goldstone mode generators to two propagators leads 
    to the time constraint $\tau\equiv t_-=t-t_+$, which leaves 
    one free time integration, to be interpreted as the relative time mismatch, $\tau$, at which the retarded and advanced time evolution traverse a common scattering path.}
    \label{fig7}
\end{figure}

The identifications $G^{+-}\sim \lambda^{-1}  B$ and $G^{-+}\sim \lambda^{-1} B^\dagger$ and Eq.~\eqref{eq:GpmPropagator} indicate that 
\begin{align}
    \label{eq:BPropagator}
    \left \langle B_{u_+ u_-} B^\dagger_{t_- t_+} \right \rangle = \Delta \delta_{\Delta t-\Delta u}\Theta\left(u- t \right),
\end{align}
i.e. a Goldstone mode `propagator', physically identical to the semiclassical pair
propagator. 
The
smallness of the scaling factor $\Delta \propto D^{-1}$ provides a posteriori
justification for the lowest order expansion of the Green function in $B$.

The above analysis gets us into a position to predict physical observables in
the ergodic regime. As an example, consider the spectral form factor
\eqref{eq:FormFactorSingle}, where Eqs.~\eqref{eq:QuasiaClassicalGreenFunction}
and \eqref{eq:PairCorrelationFunction} enter, and $\langle \dots
\rangle_\textrm{c}$ is the connected part of the ensemble average. Using
Eq.~\eqref{eq:GinBExpansion}, this becomes
\begin{align}
    \label{eq:FormFactorSingleB}
    K(t)&\sim \rho^2 \langle (BB^\dagger)_{t,0} (B^\dagger B)_{0,t} \rangle_\textrm{c}=\cr 
    & \sim \rho^2 \int_{t_+,t_-}\langle (B_{tt_-} B^\dagger_{t_-0}) (B^\dagger_{0 t_+}B_{t_+ t} ) \rangle_\textrm{c}.
\end{align}
Referring to Fig.~\ref{fig7} for a visualization, we
finally use Eq.~\eqref{eq:BPropagator} for the $B$-propagators to arrive at
\begin{align*}
    K(t)\sim \int_0^t d\tau =t,
\end{align*}
where we noted that the combination of $\delta$ and $\Theta$-functions leaves
the integral over one time coordinate $\tau=t_-=t-t_+$ unconstrained. Physically,
this time is the temporal offset between retarded and advanced pair
propagation along identical scattering paths. Irrespective of this mismatch, the
accumulated dynamical phases cancel out, and the summation over all these
contributions yields the familiar linear `ramp'. Note that this result follows
without calculation and irrespective of the detailed disorder model from symmetry and
semiclassical arguments. (For the fixation of numerical factors, we refer to the
quantitative analysis below.)

\subsection{Qudit networks}

With the above description of the single qudit ergodic phase in place, we next
discuss the influence of inter-qudit correlations. Consider neighboring qudits $j$ with
qudit states $\ket{\mu},\ket{\rho},\dots$ and $k$ with $\ket{\nu},\ket{\sigma},\dots$,
(from here on, we aim to be consistent in labelling variables pertaining to neighboring sites
by adjacent letters of the Greek or Roman alphabet)  each described by a low energy
effective theory as discussed above, and coupled by an interaction
$V=\{V_{\mu \nu,\rho \sigma}\}=\exp(i\delta H)$ where $\delta$ is a time-like parameter chosen
infinitesimally/finite in the description of continuous/discrete time dynamics,
and a Gaussian distributed generator 
\begin{align}
    \label{eq:InteractionVariance}
    \langle |H_{\mu \nu,\rho\sigma }|^2 \rangle \equiv \frac{\Lambda^2}{2D^2},
\end{align}
with adjustable strength $\Lambda$. (All other products of matrix elements of $H$ vanish
upon averaging.)

\begin{figure}[h]
    \centering
    \includegraphics[width=1\linewidth]{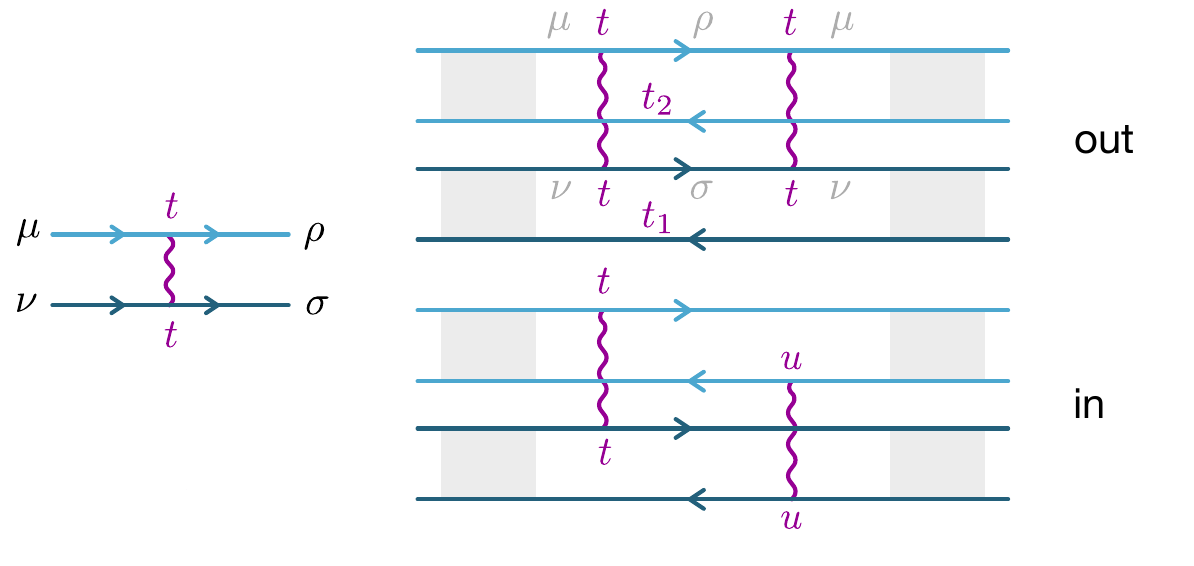}
    \caption{Left: interaction vertex coupling the quantum states of two subsystems, here distinguished by differently hued line colors. Note that a physical interaction couples amplitudes of identical causality, retarded, or advanced, as indicated by the arrows.}
    \label{fig8}
\end{figure}

\subsubsection{Synchronization dynamics}

These interactions are
instantaneous in time, and they  couple wave function amplitudes of
identical causality, retarded, or advanced (see
Fig.~\ref{fig8}, left.) Due to their Gaussian statistics, they must occur
pairwise, coupling qudits amplitudes as indicated in
Fig.~\ref{fig8}, right.  To leading order in $D^{-1}$, these
are the dominant correlation processes. Formally,  the 
Gaussian statistics penalizes  semiclassical
$B$-propagators (indicated by shading) between statistically correlated
interaction
events by lost index summations. Physically, this suppression represents the loss of
correlation under ergodic semiclassical propagation.

Specifically,  the retarded \emph{out} process shown on top is a fast succession of two
interaction events 
between  forward propagating states at time $t$ (the corresponding
advanced analog is not shown). The  times $t_1$ and $t_2$ of the
two backward propagating states in the concerned semiclassical modes are 
arbitrary. Physically, this second order
 interaction process describes the mutually decohering influence of
the two subsystems onto each other: from the perspective of system 1, system 2
is an `environment' and coherence is lost via the   \emph{out} scattering
process. As detailed below, the summation over these processes leads to an
exponential damping of the pair of interacting semiclassical modes, at a golden rule rate
$\Gamma \sim \mathrm{var}(V)\rho_2$, where $\rho_2\sim D^2$ is the 
density of states of the two-qudit Hilbert space. 

Now consider the \emph{in} process, a pair of interaction events one between two
retarded, the other between two advanced amplitudes, as indicated at the bottom
of the figure. The times of these two interactions,  $t$ and $u$, can be macroscopically
different. However, unlike with the \emph{out} process, where $t_2-t_1$ remained
unconstrained, the fixation of $t$ and $u$ removes all freedom in the relative
propagation times of the two subsystems. As a consequence, the \emph{in} process
contributes only to two-subsystem quantum amplitudes for which the
corresponding  pair propagators are dynamically  \emph{synchronized} to realize
these time constraints. 

The \emph{in} process competes with the \emph{out} process
in that it enters with opposite sign. (We label these two competitors by
\emph{in} and \emph{out} because terms of this structure feature as in and out
terms in the kinetic, or quantum master equation~\cite{Kamenev_2011} describing energy resolved
quantum transport.) The two processes cancel for the subset of synchronous
particle-hole modes, implying   that these do not suffer
dynamical damping. Physically, these  modes describe the joint
propagation of two-state amplitudes $(\psi^+_{1t}\psi^+_{2t})$ and
$(\psi^+_{1u}\psi^+_{2u})$ in the two-body chaotic Hilbert space, i.e.
correlation by interaction collapses a system described by `independent'
quasi-particle degrees of freedom $\psi^+_{1t_1}$ and $\psi^+_{2t_2}$ to one
described by single two-body amplitudes $(\psi^+_{1t}\psi^+_{2t})$. 

The iteration of this construction leads to a system-wide synchronization, with
propagating $L$-particle modes $\psi_t^+=(\psi^+_{1t}\dots \psi^+_{Lt})$ and  
$\psi^-_{u}=(\psi^-_{1u}\dots\psi^-_{Lu})$. The single relative  time
coordinate $t-u$ remains unconstrained and  enters the
description of the $L$-body ergodic mode. 

\subsubsection{Symmetries vs. interactions}
\label{sec:SymmetriesVsInteractions}

We conclude this Section with a discussion of  symmetries in the interacting
context. This point will become important when we turn to the non-perturbative
physics of thermalization. It can be skipped at first reading.

The causal symmetry breaking principle discussed in Section \ref{eq:EssentialsSymmetries} 
applies to any quantum system in a late time ergodic phase. One may even
consider it as a \emph{definition} of quantum ergodicity. However,  it is not
easily exposed in  many-body theory, with
practical consequences for the description of quantum chaos at time scales
approaching the Heisenberg time. To understand the problem, consider a
Hamiltonian describing the correlation of just two qudits $A$ and $B$, $H=H_A
\otimes \mathds{1}+ \mathds{1}\otimes H_B + H_{AB}$. Suppose we wanted to describe
this system in a path or coherent state field integral
language~\cite{Altland2023}. We then have the choice between two complementary
approaches, each highlighting different aspects of the problem. 

The first would  consider $H=\{H_{\mu \nu, \rho \sigma}\}$ as a matrix in
the space of two-qudit states $\ket{\mu} \otimes \ket{\nu}$. Defining (fermionic) annihilation
operators $\{C_{\mu \nu}\}$ of these states, one may represent the Hamiltonian
as $H=C^\dagger_{\mu \nu}H_{\mu \nu, \rho \sigma} C_{\rho \sigma}$. The path
integral action in this representation  reads 
\begin{align*}
    S[\Psi]=\int dt\, \bar \Psi (i \delta \tau_3+\partial_t-H)\Psi,
\end{align*} 
where $\Psi_{\mu \nu}\leftrightarrow  C_{\mu \nu}$ are Grassmann variables
representing the two-state operators. A state doubling    
$\Psi=(\Psi^+,\Psi^-)$ accommodates branches of retarded and advanced evolution,
distinguished by an infinitesimal shift $\pm i\delta$. In this language, the
causal symmetry is represented as $\Psi \to T \Psi$, where $T\in \mathrm{U}(2t)
$ are the symmetry transformations discussed in section \ref{eq:EssentialsSymmetries}. 
Ensemble averaging  causes symmetry breaking, the $Q$-matrices representing
its Goldstone modes.  

However, in \emph{many-body} quantum chaos, it is  more natural to
describe a Hamiltonian in a quasi-particle operator formulation: $H=
a^\dagger_\mu H_{A \mu \rho} a_\rho + b^\dagger_\nu H_{B \nu \sigma} b_\sigma +
a^\dagger_\mu b^\dagger_\nu H_{AB \mu \nu,\rho \sigma}b_\sigma a_\rho$, leading
to a path integral with action 
\begin{align}
    \label{eq:TwoBodyActionSchematic}
   S[\Psi]&=\int dt\, \big(\bar \chi (i \delta \tau_3+\partial_t-H_A)\chi +\bar \psi (i \delta \tau_3+\partial_t-H_B)\psi+ \nonumber \\
   &\qquad\qquad  + \bar \chi \bar \psi H_{AB} \chi \psi\big).  
\end{align} 
Although our symmetry principle  must 
still be present and physically effective, it is no longer visible in this representation. The origin of the problem
lies in the application of single particle coherent states
in the construction of the many-body path integral. Tensor
products of these states define separable many-body quantum states, i.e. states
void of entanglement. It has been noticed
before~\cite{tyagiSpacetimePathIntegrals2021,greenFeynmanPathIntegrals2016} that
the representation of state histories in a basis of non-entangled states ---
which is technically possible due to the over-completeness of coherent states ---
obstructs the description of strongly correlated phases. We here face this
problem in the context of the  
ergodic phase of quantum chaos. 

However, the alternative formulation in terms of two-body states outlined above, too, has its
issues: In the separable limit $H_{AB}=0$, there are \emph{two} copies of causal
symmetries, individually realized in the sectors $A$ and $B$.
Eq.~\eqref{eq:TwoBodyActionSchematic} implements these symmetries via continuous transformations $\chi \to T
\chi$ and $\psi \to U\psi $  with independent $T$ and $U$. However, the symmetry
operation $T\otimes U$ is  not represented in a manifest way in the
language of two-state variables $\Psi=\{\Psi^{a}_{\mu \nu}\}$ containing only a
single causal index. Again, we have the situation that a symmetry is present and
physical (it leads to the factorization of the  form factor
$K(t)=t^2$ in the non-interacting limit), but not  manifestly so.

We conclude that the starting point of either approach misses out on crucial
symmetries. The best we can do is choose one of them and aim
to make all symmetries emergent at some stage of the construction. Starting with
the second quantized formulation, this will be our strategy in the rest of the
paper. 

\section{Single qudit}
\label{sec:SingleQudit}
In this section, we introduce the path integral formalism for  a single qudit evolving under repeated
application of a Haar-distributed Floquet operator $U^n$. As a test observable,
we will consider its form factor $K(t)= \langle | \tr(U^t) |^2 \rangle_U $ with
the known result \eqref{eq:HaakeFormFactor}. The description of interacting
systems will then only require gradual adjustments of the formalism introduced
here.  

\subsection{Path integral}
We begin with the formulation of a real time path integral effectively
trotterizing the
evolution $|\tr(U^t)|^2=\tr(U U \dots U)\tr(U^\dagger U^\dagger \dots U^\dagger )$ in terms of $2t$ resolutions of unity between
consecutive Floquet operators $U$ ($a=+$) and  adjoints $U^\dagger$
($a=-$).  
Referring to Appendix~\ref{app:GrassmannPathIntegral} for details, we achieve this by introducing  a set of Grassmann variables
$\{\psi^{a}_{\mu
n }\}$ (which in the   framework of the coherent state path integral~\cite{Altland2023}
would represent  annihilation operators of states $\ket{\mu}$  evolving forward
or backward in discrete time.) Next define the `partition sum'
\begin{align}
    \label{eq:SingleQuditPartitionSum}
    1=Z\equiv &\int D \psi  \; e^{-S[U,\psi ]} ,\nonumber\\
&S[U,\psi]=\bar \psi^+(1-T_- U)\psi^++\bar \psi^-(1-T_+ U^\dagger)\psi^-,
\end{align}   
where $\int D \psi \equiv  \prod^{a}_{ \mu n } \int d \bar{\psi}^a_{ \mu n
}d\psi^a_{ \mu n}$, all indices are summed over, and $T_\pm$ are one-time step
translation operators, $(T_\pm f)_n=f_{n\pm 1}$. Noting that the operator $1-T_-U$ has $t$-dimensional unit
matrices on the diagonal, and $U$ on the next leading diagonal we conclude that
it has unit-determinant, explaining the normalization $Z=1$. Introducing
pre-exponential source terms, we obtain the SFF as 
\begin{align*}
    K(t)=\int D \psi \int DU \; e^{-S[U,\psi ]} (\psi^+_{t \mu}\bar{\psi}^+_{0 \mu}) (\psi^-_{0 \rho}\bar{\psi}^-_{t \rho}),
\end{align*} 
conveniently without the need for normalizing denominators. To understand this
relation without reference to its construction, just note that the Gaussian
integral over $\psi^+$ yields the matrix element $\braket{\mu t|(1-T_-
U)^{-1}|\mu 0}$. A
straightforward matrix inversion gives $\braket{\mu|U^t |\mu}$ as required.

\subsection{Haar average}

The next construction step is the  integral over the Haar distribution of the
matrices $U$. The result can  be  expressed as the integral transform 
\begin{align}
    \label{eq:ColorFlavorRealTime}
        &\int dU \, 
    e^{\bar \phi^+ U\psi^+ + \bar \psi^- U^\dagger \phi^-}=\\
    &\qquad =\int dB\, e^{-D \,\tr\ln(1+B B^\dagger)-\bar \phi^+ B \phi^- + \bar\psi^- B^\dagger \psi^+}.\nonumber
\end{align}
Eq.~\eqref{eq:ColorFlavorRealTime} is a variant of the so-called color-flavor
transform (CFT), originally developed for
supersymmetric matrix integrals~\cite{zirnbauerSupersymmetrySystemsUnitary1996}.
In it,   $B=\{B_{mn}\}$ are complex $t$-dimensional  matrices, and 
the  measure 
    $dB=\prod_{mn} dB_{mn} d\bar B_{mn} \det(1+BB^\dagger)^{-2t}$, 
is normalized in such a way that for 
$\psi,\phi=0$ the integral on the right-hand side measures the volume of the
unitary group, $\int dU$. The measure reflects the geometric interpretation of
the $B$-integral as one over the symmetric space $\mathrm{U}(2t)/\mathrm{U}(t)\times
\mathrm{U}(t)$.  However, in computing $B$-integrals we can often 
forget about its  presence. The reason is that  wide classes of
observables, among them the form factor, enjoy the mathematical property of
semiclassical exactness. In practical terms, this means that they can be computed by saddle point integration over
the stationary points of the action \emph{excluding} the
measure~\cite{andreevSpectralStatisticsRandom1995}. 
\begin{figure}[h]
    \centering
    \includegraphics[width=0.5\linewidth]{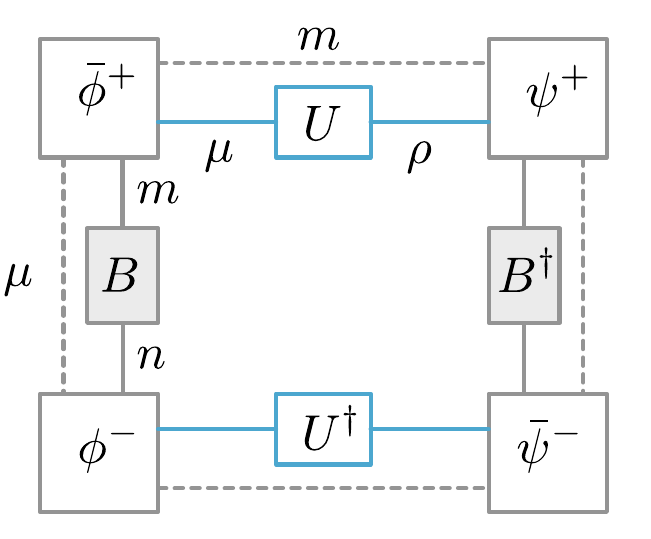}
    \caption{Color-flavor transform visualized: Bilinears of field variables 
    transforming  under $U$'s acting on their Hilbert space (`color') indices, 
    and contracted over the internal (`flavor') indices get mapped onto dual 
    bilinears where the role of color and flavor indices is exchanged, 
    flavor matrices $B$ assuming the role of the $U$'s.}
    \label{fig9}
\end{figure}

The meaning of the CFT is visualized in Fig.~\ref{fig9} in a
tensor network representation. Before the transform the field variables are
contracted through the horizontal links: they carry a nontrivial
representation of the $U$'s in  Hilbert space, indexed by the $\mu$ `color'
indices. Their internal $m$ `flavor' indices are contracted trivially.
The transform reverses  the role of the bonds: a trivial singlet
representation in color space in exchange for a nontrivial one in flavor space,
where the $B$-flavor matrices assume the role of the $U$-color matrices.

With the identifications $\bar\phi^+ = \bar \psi^+T_-$ and 
$\phi^-= T_+\psi^- $, the application of the CFT gets
us to the representation
\begin{align}
    \label{eq:CFTSingleQudit}
    Z=&\int DB D \psi\, e^{-S[B,\psi]},\nonumber\\
    &S[B,\psi]=
    D\,
    \tr\ln(1+B B^\dagger)+   \bar\psi  \underbrace{\begin{pmatrix}
       1 & B_T\\ -B^\dagger&1 
    \end{pmatrix}}_{\mathcal{G}^{-1}[B]} 
     \psi,
\end{align} 
where $\bar \psi=(\psi^+,\psi^-)$ and 
\begin{align*}
    (B_T)_{mn} \equiv (T_- B T_+)_{mn}=B_{(m-1)(n-1)}.
\end{align*} 
The merit of this transformation is that it trades the average over
high-dimensional random color fields, $U$, for the simpler integration over
flavor degrees of freedom, where the denotation `$B$' indicates that the flavor
fields will assume the role of the  Goldstone mode generators discussed in
section~\ref{sec:Essentials}. To understand this point, execute the Gaussian
$\psi$-integral to obtain
\begin{align}
    \label{eq:ActionFunctionalBSingle}
    Z=&\int DB \, e^{-S[B,\psi]},\nonumber\\
    &S[B,\psi]=
    D\,
    \tr\ln(1+B B^\dagger)-D\,
    \tr\ln(1+B_T B^\dagger).
\end{align} 
This representation demonstrates that fields $B_T\simeq B$ with slow time
dependence are soft modes with vanishing action. In other words, they are the
Goldstone modes of our symmetry breaking phenomenon. 

\subsection{Luttinger-Ward functional}

However, to prepare for the subsequent extension to the interacting context, we
take the  detour to not integrate over the $\psi$'s right away. Instead, we first
introduce the quasiclassical Green functions $G$ as integration variables to
obtain an integral representation interchangeably known as
Luttinger-Ward~\cite{luttingerGroundStateEnergyManyFermion1960}, or $G
\Sigma$-functional~\cite{rosenhausIntroductionSYKModel2019}. We wish to enforce
the identification
    \begin{align}
    \label{eq:PsiGLockingDiscrete}
    -\frac{ia }{D}  \sum_\mu \psi^a_{\mu m}\bar \psi^{b}_{\mu n}\equiv  G^{ab}_{mn},   
\end{align}
i.e. Eq.~\eqref{eq:QuasiaClassicalGreenFunction}, for discrete time indices. 
To this end, we introduce a Lagrange multiplier, 
\begin{align}
    \label{eq:LagrangeMultiplier}
    S[B,\psi]\to S[B,G]+ D\,\tr\left( \Sigma\left(  G- \frac{i}{D}   \tau_3 (\psi \bar \psi) \right)\right),
\end{align}
where $\Sigma = \{\Sigma_{mn}^{ab}\}$ are matrices sharing the index
structure of $G$. The quadratic integration over $\psi$ then produces the `$G \Sigma$' action
    \begin{align}
    \label{eq:GSigmaActionSingle}
     S[G,\Sigma,B]&=
          D\,  \big(\tr\ln(1+B B^\dagger)
         -\nonumber \\
         &\tr \ln \left( \mathcal{G}^{-1}(B)-i \Sigma \tau_3  \right)  
     + 
    \tr(\Sigma G)\big).
\end{align}
In the $G \Sigma$-representation, the SFF assumes the form 
\begin{align*}
    K(t)=D^2\left\langle G_{t1}^{++}G^{--}_{1t} \right\rangle_{B,G,\Sigma}. 
\end{align*} 
where $t=1$ is set to the starting time step for the discrete model.

\subsection{Stationary phase analysis}

To make progress with this expression, we turn to a stationary phase analysis,
stabilized by the largeness of  $D$. Variation of the action yields the saddle
point equations
\begin{align}
    \label{eq:StationaryPhaseSingle}
    0&=\delta_{\Sigma} S[\bar G,\bar \Sigma,B]
    =
    D\left( \bar G + i  \tau_3 (\mathcal{G}^{-1}(B)-i\bar \Sigma\tau_3)^{-1} \right),\nonumber \\
    0&=\delta_{G}S[\bar G,\bar \Sigma,B]=D\bar \Sigma=0.
\end{align} 
Their solutions 
\begin{align}
    \label{eq:SimplifiedStationaryPhase}
\bar \Sigma=0,\qquad  \bar G = -i \tau_3 \mathcal{G}(B),
\end{align} 
have a status similar to the self-consistent Born equations describing  single
Green functions in the presence of hermitian disorder. In that case, averaging
over configurations generates  an effective self energy describing the
attenuation of an averaged propagator due to impurity scattering. However,  for
random unitary operators, the vanishing of all moments $\langle U^m \dots
\rangle_U=0$ implies the absence of average self energies, i.e. the first of the
two equations~\eqref{eq:StationaryPhaseSingle}. The second equation is
solved by inversion of the block matrix $\mathcal{G}^{-1}$ defined in
Eq.~\eqref{eq:CFTSingleQudit} (e.g., by geometric series summation): 
\begin{align}
    \label{eq:MeanFieldGSingle}
    \bar G =  \frac{i}{2}(Q_T-\tau_3), 
\end{align} 
where the matrix $Q_{T}$ is defined in Eq.~\eqref{eq:QMatrixB}, except that the
matrix $B\to B_T$ is to be replaced by the one-step time translated $B_T$. (The
matrix $B^\dagger$ remains as it is.) 

At this point, we have established that the  Green function
is governed by the Goldstone modes introduced in
Section~\ref{sec:Essentials}. Comparing to the previously discussed matching
$G=i\lambda Q$, the presence of the constant matrix $\tau_3$ in
Eq.~\eqref{eq:MeanFieldGSingle}, 
and the absence of a dimensionful
energy scale $\lambda$, are consequences of the discrete time definition of the
present model.

\subsection{Fluctuation action}

We next need to construct a fluctuation action describing the long time dynamics
of our system. In principle, the term `fluctuations' refers to deviations
$(\delta \Sigma,\delta G)$ around the mean field $(0,\bar G)$, as well as  to
fluctuations of $B$ on the saddle point manifold parameterized by $\bar G$. It
is straightforward to see that fluctuations of the former type do not exist in
the present simple model. (Formally, this follows from the fact that the action
\eqref{eq:GSigmaActionSingle}  
contains $G$ only in linear order.) Turning to the more
interesting $B$-fluctuations, the substitution of $\mathcal{G}^{-1}(B)$ into the
`tr ln' leads to Eq.~\eqref{eq:ActionFunctionalBSingle}, i.e. a representation
of the functional which a straightforward integration over $\psi$ would have
produced without the need for a stationary phase program. (However, the merit of
the $G \Sigma$-representation will become evident once we turn to interactions.)

\subsubsection{Nonlinear $\sigma$-model }
Introducing the discrete derivative in the center time dependence, 
\begin{align*}
 dB_{nm}\equiv (B-B_T)_{nm}=B_{nm}-B_{(n-1)(m-1)},
\end{align*}
the next step is an expansion of the logarithmic action to leading order in this
difference: 
\begin{align}
    \label{eq:S0DiscreteTime}
    &S_0[B]\equiv D\, \tr\left( \frac{1}{1+B^\dagger B}B^\dagger d B \right).
\end{align}
To establish contact with the literature, we note that a straightforward
reordering of terms (always to leading order in the difference operator $d$),
brings the action into the form
\begin{align}
    \label{eq:SigmaModelInvariant}
    S_0[T]=\frac{D}{2}\tr(dT \tau_3 T^{-1}),
\end{align} 
which is a discrete time version of the nonlinear $\sigma
$-model~\cite{Kamenev_2011}.

While Eq.~\eqref{eq:S0DiscreteTime} will be the
representation primarily used in our further construction of the theory,
Eq.~\eqref{eq:SigmaModelInvariant} has the advantage of making the symmetries of
the model manifest: Transformations $T\to T_0 T$, where
$(T_0)_{nm}=T_0 \delta_{nm}$ does not depend on time indices, leave the action
invariant. A closer analysis~\cite{altlandWignerDysonStatisticsKeldysh2000}
reveals the existence of families of saddle points besides $Q=\tau_3$ (i.e.
configurations of the $Q$-matrix around which the expansion in fluctuations
starts at quadratic order) generalizing the concept of
the so-called Altshuler-Andreev (AA) saddle
points~\cite{andreevSpectralStatisticsRandom1995} to the real time framework.
Gaussian integration over these saddles generates the `cutoffs' at $t=D$
terminating the linear increase of the form factor
Eq.~\eqref{eq:FormFactorDiscreteTime}. However, in this paper, we will take the shortcut
to mostly work at $t<D$, where the AA saddles do not contribute, and only point
out their contribution in final results. 

\subsubsection{Quadratic action}

To establish contact to the semiclassical principles discussed in section
\ref{sec:Essentials}, consider the quadratic expansion, 
\begin{align}
    \label{eq:S0DiscreteTimeQuadratic}
    S_0^{(2)}[B]\equiv  D\, \tr\left(B^\dagger d B \right),
\end{align}
This Gaussian weight  defines
the discrete time propagator of the theory, 
\begin{align}
    \label{eq:BContractionDiscrete}
    \langle B_{n_+,n_-} B^\dagger_{m_- ,m_+} \rangle \equiv \Pi_{n,m},
\end{align} 
as the inverse  
\begin{align*}
    &D(d\Pi)_{n,m}=D (\Pi_{n,m}-\Pi_{n-1,m})= \delta_{n,m},
\end{align*}
where $n\equiv(n_+,n_-)$, $n-1=(n_+-1,n_--1)$, and $\delta_{n,m}=\delta_{n_+,m_+}\delta_{n_-,m_-}$. It is straightforward to check, that this relation is satisfied by the solution 
\begin{align}
    \label{eq:BPropagatorDiscrete}
    \Pi_{n,m}=\frac{1}{D}\delta_{\Delta n,\Delta m}\Theta(\bar n-\bar m),
\end{align}
where $\Delta n = n_+-n_-$ and $\bar n=(n_+ + n_-)/2$, and likewise for the
$m$-indices. $\Theta(0)=1$ is used. In other words, $\Pi$ is the discrete time version of the
semiclassical ergodic propagator Eq.~\eqref{eq:BPropagator}. 

\subsection{Spectral form factor}
\label{sec:SFFSingle}

The computation of physical observables now proceeds along the lines of a
straightforward protocol: i) express their coherent state $\psi$-representation
through the quasiclassical Green function, ii) expand the latter to any required
order in $B$-fluctuations (including saddle points beyond the standard saddle
$G=-i\tau_3$ if nonperturbative information is required) iii) do the
$B$-integrals. 

Considering the spectral form factor as an example, step i) leads to the
representation Eq.~\eqref{eq:FormFactorSingle}, and ii) to the discrete time
analog of Eq.~\eqref{eq:FormFactorSingleB}, 
\begin{align}
    \label{eq:FormFactorBSingle}
    K(t)=D^2 \left\langle (BB^\dagger)_{t1}(B^\dagger B)_{1t} \right\rangle. 
\end{align} 
With the matrix products resolved as $(B B^\dagger )_{t1}=B_{tn}B^\dagger_{n1}$,
the Gaussian contraction according to Eq.~\eqref{eq:BContractionDiscrete} yields 
\begin{align*}
    K(t)= D^2\sum \limits_{n,m}\Pi_{tn,1m}\Pi_{mt,n1}=t.
\end{align*} 
As mentioned above, the extension to times $t>D$ requires integration over
non-standard saddle points. For the execution of this program in the Fourier
conjugate energy representation we refer to
Ref.~\cite{altlandWignerDysonStatisticsKeldysh2000} and for the adaption of this
approach to the time domain to a forthcoming publication.

\section{Brickwall model}
\label{sec:CircuitNetworks}

\begin{figure}[h]
    \centering
    \includegraphics[width=0.75\linewidth]{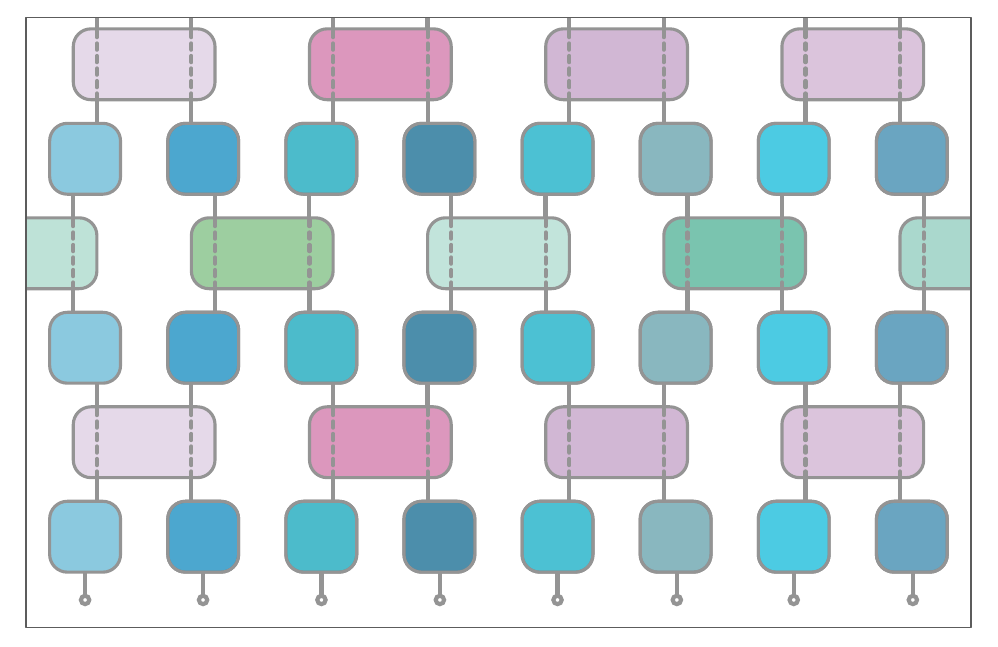}
    \caption{A brickwall design describing the alternating application of random one- and two-qudit unitary operators. Statistically independent  realizations are indicated by color coding. Note the repeated application of identical operators in vertical (time) direction, defining a four-step Floquet dynamics. The limiting case of absent correlations is indicated by dashed lines. }
    \label{fig10}
\end{figure}

With the path integral  of the single qudit  in
place, the extension to the network shown in
Fig.~\ref{fig10} is straightforward. There, the blue boxes
define an array of $L$  qudits of dimension $D$, governed by statistically
independent  unitaries, each  repeated periodically in
vertical discrete time. Neighboring qudits are coupled by tensor products of two
qudit unitaries $V={V_{\mu  \rho,\rho\sigma}}$, where $V=\exp(i H)$ and
$H$ distributed according to \eqref{eq:InteractionVariance} (with dimensionless $\Lambda$).

We label each product $U V_{\mathrm{pink/green}}$, by discrete time
steps $m$, with  pink/green interactions for even/odd $m$. The  two stages of
each step are represented by a `stage index' $c=\pm 1$, where the passage from
$(m-1,+1)$ to $(m,-1)$ is governed by the tensor product of single qudit
operators $U$, and that from $(m,-1)$ to $(m,+1)$ by interactions, $V$. On the
retarded contour the sequence is traversed in reverse order,  
\begin{alignat*}{9}
    &+&\qquad \dots \to & (m-1,+1)&\stackrel{U}{\longrightarrow} &(m,-1)& \stackrel{V}{\longrightarrow} & (m,+1) &\to \dots &\cr 
    &-&\qquad \dots \leftarrow & (m-1,+1)&\stackrel{U^\dagger}{\longleftarrow} &(m,-1) &\stackrel{V^\dagger}{\longleftarrow} 
    & (m,+1) &\leftarrow \dots&. 
\end{alignat*} 

\subsection{Path integral}

We obtain the path integral for this protocol by straightforward
generalization of the previous construction. Apart from site indices
$j=1,\dots,L$ and the stage index $c$, it contains  the two-qudit interaction
as a new element. With Grassmann variables $\psi^{a}_{j \mu m c } $, its
unit-normalized partition sum assumes the form $Z=\int D \psi \exp(-S[\psi])$,  
where the action is given by 
\begin{align}
    \label{eq:SMicroscopic}
    &S[\psi]=\bar \psi \psi +S[U,\psi]+S[V,\psi],\\
&\quad S[U,\psi]=\bar\psi^+_{j-1} T_- U_j \psi^+_{j1}+\bar\psi^-_{j1} U^\dagger_j T_+ \psi^-_{j-1},\nonumber\\
&\quad S[V,\psi]= -i\bar\psi^a_{a}\psi^a_{-a}- \bar \psi^a_{ja}\bar \psi^a_{ka}
\left(- i a H +\frac{1}{2}H^2 \right)_{jk} \psi^a_{k-a}\psi^a_{j-a}.\nonumber
\end{align}
Referring to Appendix~\ref{app:GrassmannPathIntegral} for details,
$\bar{\psi}\psi \equiv \sum \bar\psi^{a}_{j \mu n c }\psi^{a}_{j \mu n c }$ is
the  Gaussian weight resulting from the insertion of Grassmann
Kronecker-$\delta$'s at each time step. The action $S[U,\psi]$ describes  single qudit
evolution at steps $(m-1,1)\rightarrow (m,-1)$, where $(T_\pm
f )_m=f_{m \pm 1}$ as before. (Note that in this action, the stage indices are
determined by the $a=\pm1$ causal index, hence the double appearance of $a$ as
super- and subscript.) Finally, $S[V,\psi]$ describes interactions $(m,-1)\to (m,1)$ with
$V=\exp
(-i  H)\simeq \mathds{1}-i  H -  H^2/2$ expanded to second
order in $  H$. The second term in $S_V$ is the `normal-ordered' quartic
Grassmann representation of the two body operators $H$ and $H^2$, where the
expansion to second order is required to describe the time-local  second order interaction processes
discussed in Section~\ref{sec:Essentials}, and in more detail below.

\subsubsection{Averaging over randomness}

Our system contains two sources of randomness, the Haar-distributed unitaries
$U$, and the Gaussian distributed $H$. Averaging over $U$  yields the
generalization of Eq.~\eqref{eq:CFTSingleQudit} to multiple sites, 
\begin{align}
    \label{eq:BPsiAction}
    &S[U,\psi] \to S[B,\psi] = D \sum_j \tr\ln(1+B_j B^\dagger_j)+ \\
    &\quad+  \sum_j \left( \bar\psi^+_{j-1}B_{Tj}  \psi^-_{j-1}- \bar\psi^-_{j1} B^\dagger_j  \psi^+_{j1}
 \right) .\nonumber
\end{align}
Averaging  over $H$ amounts to a
straightforward Gaussian integral with variance \eqref{eq:InteractionVariance}.
The result reads
\begin{align}
        \label{eq:PsiInteractionAction}
S[V,\psi]&\to S_\textrm{int}[\psi]\equiv S_{\textrm{o}}[\psi]+ S_{\textrm{i}}[\psi],\cr
&S_{\textrm{o}}[\psi]=\frac{\Gamma}{2}\sum_{\langle j,k \rangle,m,a}
((\psi\bar \psi)_k\odot (\psi \bar \psi)_j)^{a,a}_{m -a,ma},\cr
&S_{\textrm{i}}[\psi]=\frac{\Gamma}{2 D^2} \sum_{\langle j,k \rangle, mn,ab}ab\, 
((\psi \bar \psi)_j\odot (\psi \bar \psi)_k)^{a,b}_{m-a,nb}\times \cr
&\hspace{2cm}\times  
((\psi \bar \psi)_j\odot (\psi \bar \psi)_k)^{b,a}_{n -b,ma},
\end{align} 
where we use the Hadamard product notation Eq.~\eqref{eq:HadamardProduct} for field bilinears 
\begin{align}
    \label{eq:ColorSinglet}
    (\psi \bar \phi)_j\equiv \psi_{j\mu} \bar \phi_{j\mu},\cr
    ((\psi \bar \phi)_j)^{a,b}_{mc,nd}\equiv \psi^a_{j\mu mc} \bar \phi^b_{j\mu nd},
\end{align}
and the sum is over pairs of nearest neighbor sites $\langle j,l \rangle$
subject to the condition that on odd/even time slices $m$ different site
pairings are being summed over. We also defined
\begin{align}
    \label{eq:GammaDef}
    \Gamma \equiv \frac{\Lambda^2}{4}.
\end{align}
Note the golden rule like structure of the interaction constant as a product of
the effective dimensionless  interaction strength, $(\Lambda)^2/2D^2$,
multiplied by the two-body density of states $D^2$ of final states.

\begin{figure}[h]
    \centering
    \includegraphics[width=.8\linewidth]{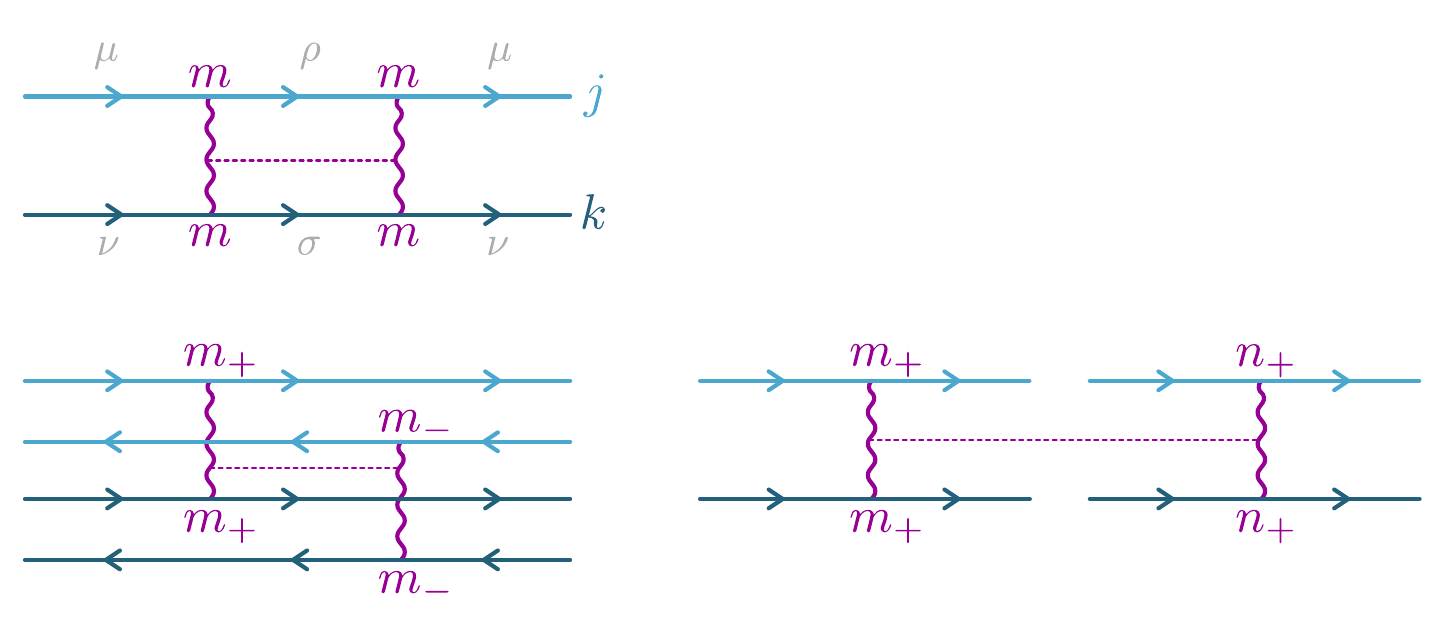}
    \caption{Top: The vertex $S_{\textrm{o}}$ in
    Eq.~\eqref{eq:PsiInteractionAction}. The dashed line indicates the
    statistical average with second moment \eqref{eq:InteractionVariance},
    enforcing a Hilbert space structure as indicated. The bottom-left block
    represents the instantaneous second order interaction event at discrete time
    $m$. Only fields of identical causality (indicated by the direction of
    arrows) contribute to this `self energy vertex'. Bottom-right: Eighth order (in
    $\psi$) vertex $S_{\textrm{i}}$ describing the correlation of scattering
    processes involving states of opposite (left) or identical (causality) at
    different discrete times as indicated. (Our subsequent analysis will show
    that the vertex shown in the bottom left does not contribute at leading
    order in the $D^{-1}$-expansion.) }
    \label{fig11}
\end{figure}

While the first contribution to the interaction, $S_{\textrm{o}}$, results from
the averaging of the exponent $\frac{1}{2}\langle \bar \psi H^2 \psi\rangle$, 
the second, $S_{\textrm{i}}$, is obtained as 
    $\frac{1}{2}\langle (\bar \psi H \psi)^2 \rangle$. 
A graphical interpretation of these two
interaction vertices is  indicated in
Fig.~\ref{fig11}, in a color coding matching that of the
earlier Fig.~\ref{fig8}. The `out' vertex
$S_{\textrm{o}}$, visualized on top, describes 
virtual two-body scattering from two qudit $(j,k)$ pair states $\ket{\mu,\nu}$
into the transient states $\ket{\rho,\sigma}$ and back, where the identity of
initial and final states is required by the Gaussian correlation of matrix
elements. This process is instantaneous in the
discrete time index $m$. By contrast, the `in' vertex represents the statistical
correlation of scattering events taking place at different times $m$, $n$.
These two scattering events may either involve propagators of opposite (left) or
identical (right) causality. The latter will turn out to be negligible to
leading order in the $D^{-1}$-expansion.

\subsubsection{Luttinger-Ward functional}

The nonlinearity of the $S_\textrm{i}$ in $\psi$ implies that we can no
longer integrate over these variables in closed form. However, at this point
the advantage of the $G\Sigma$-construction becomes evident: the interaction
contains $\psi$ in the form of color singlets Eq.~\eqref{eq:ColorSinglet}.  Lagrange multipliers
Eq.~\eqref{eq:LagrangeMultiplier} implemented locally at each site allow us to trade
all of these for quasiclassical Green
functions~\eqref{eq:PsiGLockingDiscrete}. The result of this replacement reads
as 
\begin{align}
    \label{eq:InteractionG}
    &S_{\textrm{o}}[G]=-\Gamma D^2\sum_{\langle j,k \rangle,n,a}(G_k\odot G_j)^{aa}_{n -a,na},\cr
    &S_{\textrm{i}}[G]=\Gamma D^2\sum_{\langle j,k \rangle, nm,ab}ab
        \times \cr
    &\hspace{1.5cm}\times  
    (G_j\odot G_k)_{m-b,na}^{ba}(G_j\odot G_k)^{ab}_{n-a,mb}.
\end{align} 
Adding to these terms the non-interacting part of the action, comprising the $B$-matrices and the $G
\Sigma$-Lagrange multiplier structure, we obtain
\begin{align}
    \label{eq:GSigmaAction}
    S[G,\Sigma,B]&=D\, \sum_j \big(\tr\ln(1+B B^\dagger)-\tr \ln \left( \mathcal{G}^{-1}(B)-i \Sigma \tau_3  \right)
    \nonumber\\
    &\qquad 
    + 
    D\,\tr(\Sigma G)\big)_j + S_\textrm{int}[G].
\end{align}
Here,  $\mathcal{G}^{-1}(B)$ is a $4\times 4$ block-matrix  in the tensor product of
causal and stage index space.  It  is implicitly defined through the quadratic
$\psi$-action in Eqs.~\eqref{eq:SMicroscopic} and \eqref{eq:BPsiAction},  and
explicitly  in Appendix~\ref{app:LuttingerWard}, Eq.~\eqref{eq:CalGDefinition}.
The straightforward inversion of this matrix leads to 
  \begin{align}
    \label{eq:CalGDefinition}
{\cal G}(B)
&=
\frac{1}{2}\left(\tau_3 Q-1\right)\otimes J_2+\textrm{const.}, 
\end{align}
where $Q(B)$ is defined in Eq.~\eqref{eq:QMatrixB} and
   $(J_2)_{cd}=1$  in stage space. The constant matrix  $\textrm{const.}=-
 \frac{1}{2}((1+\tau_3)\otimes \sigma^+ + (1-\tau_3)\otimes  \sigma^-)$ with
 $\sigma^\pm=\frac{1}{2}(\sigma_1 \pm i \sigma_2)$
 in stage space will not feature in our further discussion.  

For later reference, we note that in microscopic path integral variables the
form factor generalized to $L$-qudits assumes the form
\begin{align*}
    K(t)\equiv \left \langle \prod_j \psi^+_{jt1} \bar \psi^+_{j1-1}   \psi^-_{j1-1} \bar \psi^-_{jt1} \right \rangle,
\end{align*}
where $c=\pm 1$ are stage indices. By design of the path integral,
this correlation function computes the discrete time evolution between states
$\ket{\mu_1}\otimes \dots \ket{\mu_L}$ and the same final state, summed over all
Hilbert space amplitudes, in retarded (first) and advanced (second factor)
order. The corresponding $G$-representation is given by
\begin{align}
    \label{eq:FormFactorDiscreteG}
    K(t)\equiv (-D^2)^L \left \langle \prod_j G^{++}_{j,t1}G^{--}_{j,1t} \right \rangle.
\end{align}
An expansion of the Green functions to
leading quadratic order in $B$-fluctuations generalizes the single qudit  Eq.~\eqref{eq:FormFactorSingleB} to 
\begin{align}
    \label{eq:FormFactorMultiSiteB}
K(t)\approx (-D^2)^L\prod_j \left \langle (B_{tn} B_{n1}^\dagger B^\dagger_{1m} B_{mt})_j \right \rangle_B,
\end{align} 
with discrete time indices independently summed for each $j$.

\subsubsection{Stationary phase analysis}

Following the same protocol as in the single qudit case, we proceed by stationary phase
analysis, where the generalization of equations~\eqref{eq:StationaryPhaseSingle} to the action
Eq.~\eqref{eq:GSigmaAction} reads as 
\begin{align*}
    0&=\delta_{\Sigma_j} S[\bar G,\bar \Sigma,B]=D\left( \bar G_j + i  \tau_3 (\mathcal{G}^{-1}(B_j)-i\bar \Sigma_j\tau_3)^{-1} \right),\cr
    0&=\delta_{G_j}S[\bar G,\bar \Sigma,B]=D\bar \Sigma_j + \delta_{G_j} S_\textrm{int}[\bar G].
\end{align*} 
At first sight, the presence of $S_\textrm{int}$ appears to significantly
complicate these equations. However, (cf. the single site solution Eq.~\eqref{eq:SimplifiedStationaryPhase})
\begin{align}
    \label{eq:StationaryPhaseArray}
\bar \Sigma=0,\qquad  \bar G_j = -i \tau_3 \mathcal{G}(B_j),
\end{align} 
at $B_j=0$  continue to be stationary configurations of the
theory. Fluctuations around this configuration start at quadratic order, as we
are going to demonstrate momentarily. The physical reason for the robustness of
this saddle point is easiest to see within the continuous time framework, and we
postpone its discussion to section \eqref{sec:HamiltonianSystems}. However,
unlike with the non-interacting theory, large $B$-flucutations away from the
semiclassical limit, $B=0$, do not trivially leave the interaction invariant.
Identifying which of these are admissible in the interacting context will be an
essential part of our analysis below. 
For the time being, we proceed by expansion around the saddle point
defined above.

\subsection{Interactions}
\label{sec:Interactions}

\begin{figure}[h]
    \centering
    \includegraphics[width=.6\linewidth]{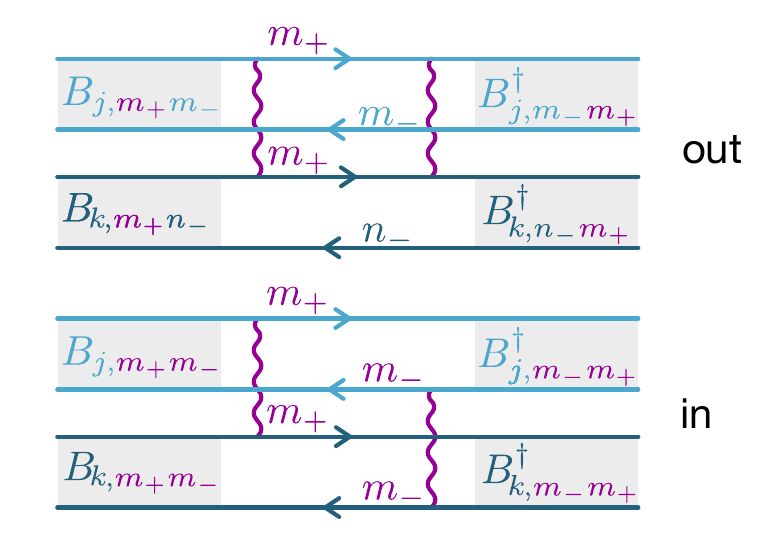
}
    \caption{The interaction vertex expanded to lowest order in the Goldstone mode generators of two neighboring sites. The locking of temporal indices enforced by the instantaneous nature of the interaction is indicated in color code. Note that the out-term (top) leaves two time variables unconstrained, while the in-term leads to full synchronization.}
    \label{fig12}
\end{figure}

In this central section we quantitatively describe  the synchronization dynamics
discussed qualitatively in section~\ref{sec:SummaryOfResults}. In the absence of
interactions, the form factor Eq.~\eqref{eq:FormFactorDiscreteG} factorizes and
the execution of $L$ decoupled path integrals yields $K(t)=(K_1(t))^L$. To
include interactions, we substitute Eq.~\eqref{eq:StationaryPhaseArray} into the
interaction Eq.~\eqref{eq:InteractionG} and expand  to leading, quartic order in
the $B$-generators. Referring to Appendix \ref{app:LuttingerWard} for details,
we obtain
\begin{align}
    \label{eq:SintLowestOrder}
    S_\textrm{int}^{(4)}[B]&= \Gamma D^2\sum_{\langle j,k \rangle,n,m}(B_{T m_+ m_-}  B^\dagger_{m_- m_+})_j 
    (B_{T n_+ n_-}B^\dagger_{n_- n_+})_k \nonumber  \\
    & \qquad \times \big( \delta_{n_+ m_+}+ \delta_{n_- m_-}-2 \delta_{n_+ m_+}\delta_{n_- m_-}\big),
\end{align}     
where the first two (the third) Kronecker-$\delta$ originate from the out-(in-)term, cf. Fig.~\ref{fig12}.

The crucially important feature of this vertex is its vanishing on temporally
synchronized configurations $(B_j\odot B_k)_{m_+ m_-}=B_{j,m_+m_-}B_{k,m_+m_-}$.
In Appendix \ref{app:InteractionBeyondQuartic}  
we show that this vanishing extends to all orders in the expansion  in
Goldstone mode generators. The conclusion of this discussion is  that 
\begin{align}
    \label{eq:XModeDefinition}
    X_{mn}=\left( \odot_{j=1}^L B_j \right)_{mn}=\prod_{j=1}^L B_{j,mn},
\end{align}
are collective field amplitudes commuting with all interaction vertices. The single generator $X=\{X_{mn}\}$ then parameterizes a
$Q$-matrix describing the ergodic phase of the coupled system. 

In the following,
we describe the crossover dynamics governing the passage from the short time
limit of decoupled form factors $K_1^L(t)$ to the late time limit with $K_L(t)$.
To this end, we focus on the interaction vertex expanded to
quartic order in $B$. Higher order contributions are suppressed in inverse powers of $D$ (cf. the discussion of Section
\ref{sec:Essentials}). The goal  is to sum over an infinite series of
vertex insertions, physically representing the competition of in and out
scattering processes discussed above. While this may look like a complicated task, the triviality of the
ergodic propagator Eq.~\eqref{eq:BPropagatorDiscrete} provides sufficient
tailwind to execute it in closed form. In concrete terms, we proceed
along a succession of four steps, executed in detail in Appendix \ref{app:HSInteraction}:
\begin{enumerate}
    \item Hubbard-Stratonovich decoupling of the interaction by introduction of
    a time-bilocal  auxiliary field, $\phi_{jn_+n_-}\equiv \phi_{jn}$. 
    \item The quadratic $B$-action including this field then reads (symbolic
    notation), $\sum_{jn}B_{jn}^\dagger(d -\phi_{j,n})B_{jn}$, with the discrete
    time derivative
    $dB=B-B_T$.
    Reminiscing a scalar potential coupling, it suggests removing  $\phi$  from the action by a time-dependent
    gauge transformation. With the  phase factor $\Theta_n=\sum_m
    \phi_{n-m}$, this is achieved by the change of variables $B_n\to
    \exp(\Theta_n) B_n$. 
    \item This transformation removes the interaction potential  from the
    action. However, it reappears in operators representing correlation
    functions in the form of gauge-phase factors containing discrete time
    integrals over $\Theta_n$.  
    \item The final Gaussian integration over these variables then 
    effectively sums over interaction processes to infinite order in
    perturbation theory. 
\end{enumerate}

\subsection{Form factor (semiclassical)}
Applied to the form factor Eq.~\eqref{eq:FormFactorMultiSiteB}, the protocol
above leads to 
\begin{align}
       \label{eq:FormFactorInHS}
       K(t)&\to D^{2L} \sum_{\{m,n\}}\prod_j \left \langle (B_{tn} B_{n1}^\dagger B^\dagger_{1m} B_{mt})_j \right \rangle_B \times \nonumber\\
       &\qquad \left \langle e^{\sum_j(\Theta_{tm}-\Theta_{1m}-\Theta_{n1}+\Theta_{nt})_j} \right \rangle_\phi, 
\end{align}
with the  abridged notation,  $(B_{tn})_j\equiv B_{jtn_j}$ or
$(\Theta_{tn})_j=\Theta_{jtn_j}$, etc. This expression exemplifies how  the
gauge phase factors factorize correlation functions into an interaction
contribution times a factor describing the ergodic dynamics of the individual
systems. As in our discussion of the non-interacting qudits (cf.
Section~\ref{sec:SFFSingle}), the role of the latter is to enforce constraints
$t-n_j=m_j$ on the intermediate time variables. The final integration over the
Hubbard-Stratonovich phases detailed in Appendix \ref{app:HSInteraction} then
leads to the result
\begin{align}
    \label{eq:FormFactorDiscreteTime}
    K(t)=\sum_{\{n\}} e^{-S[n]},\qquad S[n]=\Gamma t \sum_{\langle j,k \rangle}(1-\delta_{n_j,n_k}),
\end{align}
obtained earlier in
Ref.~\cite{chanSpectralStatisticsSpatially2018} for systems of qudits correlated
by diagonal matrices of randomly fluctuating phases. It was there interpreted
as the partition sum of a $t$-state Potts model, i.e. a system of $L$
fictitious spins, $n_j$ of size $t$, subject to nearest neighbor interaction
$S[n]$. In this reading, the $t$ configurations of the fully
synchronous/magnetized state $n=n_1=\dots=n_L$ with action $S[n]=0$ contribute the
long time asymptotic $K(t)\to t$. More generally, the  summation
over all configurations yields 
\begin{align}
    K(t)&=  \left( 1 +\left( t-1 \right)e^{-\Gamma t}\right)^L + \left( t-1 \right)\left( 1-e^{-\Gamma t}  \right)^L, 
\end{align}
which for $t\gg1$ simplifies to Eq.~\eqref{eq:FormFactorPotts},
interpolating between  $K(t)\approx t^L$ for $\Gamma t\ll 1$ and
$K(t)\approx t$ for $\Gamma t\gg 1$.

\subsection{Form factor (non-perturbative)}

Eq.~\eqref{eq:FormFactorPotts} is obtained in lowest order perturbation theory
in the $D^{-1}$-expansion of the theory. 
There are two time scales at which these results need
refinement. The first are times $t\sim D$ comparable to the Heisenberg time of
individual qudits, where their  dynamics is no longer adequately
described by lowest order expansion in the qudit pair propagators
Eq.~\eqref{eq:BPropagatorDiscrete}. Both, in applications and numerical
simulations, $D$ need not be particularly large (we use $D=12$ in the largest
systems $L=5$ of our numerical analysis), implying that substantial corrections
at $t\sim D$ time scales can be practically relevant. Below, we will reason that
the extension of the analysis to $t\sim D$ non-perturbative fluctuations requires
a replacement $t \to K_1$  
in the first bracket on the right-hand side of
Eq.~\eqref{eq:FormFactorPotts}.   

The second is the much larger Heisenberg time of the full system, $t\sim D^L$.
This scale corresponds to the smallest energy of the system, its many-body level
spacing, and for sizeable interaction scales, is embedded deeply into the
ergodic regime, $D^L\gg \Gamma^{-1}$. The generalization of the analysis to
non-perturbative fluctuations at this timescale amounts to a replacement $t\to
K_L(t)$ 
in the second bracket at the right-hand side of
Eq.~\eqref{eq:FormFactorPotts}, so that we arrive at the result
Eq.~\eqref{eq:FormFactorDiscreteNonPert}. In view of the fact that this
generalization of Eq.~\eqref{eq:FormFactorPotts} is relatively obvious (but
requires substantially more computationally overhead), readers primarily
interested in the generalization to other system classes are invited to proceed
to Section \ref{sec:Symmetries}. For all others, the following two subsections
outline the non-perturbative extension of the so-far analysis. The full
technical execution of this program, with a focus on the regime of ultra-weak
interactions $\Gamma^{-1}\lesssim D^{-1} $, where mechanisms of Fock- or
many-body localization begin to play a role, will be the subject of a
forthcoming publication.    

\subsubsection{Time scales $\sim D$ }

At the time $t=D$, the form factor $K_1(t)$ of a single qudit levels off at the
value $K_1(t)=D$. The representation $K_1(t)=t - (t-D)\Theta(t-D)$, indicates
that the termination of the semiclassical ramp $K_{1,\textrm{semicl.}}=t$   is
due to contributions to the path integral non-vanishing at times $t>D$.  These
terms originate in the so-called Altshuler-Andreev (AA)
saddles~\cite{andreevSpectralStatisticsRandom1995}, stationary points of the
path integral different from the so-called `standard saddle' $Q=\tau_3\otimes
\mathds{1}$, where $\mathds{1}$ is the $t$-dimensional     unit-matrix in
discrete time space. Referring to
Ref.~\cite{altlandWignerDysonStatisticsKeldysh2000} for a quantitative analysis
of the form factor with inclusion of these stationary points, we here just
mention a few aspects of relevance to our present analysis.

The AA saddle points have to do with the  action cost associated with the time
translation operator in the color-flavor transformed action
\eqref{eq:S0DiscreteTime}. If it were absent, all  $Q$-matrices $Q=T
(\tau_3\otimes \mathds{1})T^{-1}$ would have vanishing action. Its presence
isolates discrete sets of stationary points on this high-dimensional fluctuation
manifold,  where the AA saddles are defined by flipping just two-entries of the
$2t$-dimensional matrix  $\tau_3\otimes \mathds{1}$. The sum over the
fluctuation integrals around all these stationary points yields the full form
factor; fluctuation integrals around more complex stationary points vanish. 

The interplay of the AA saddles with the interaction vertex is a more
complicated which we have not investigated in detail. However, their proximity
to the standard saddle indicates that they will be subject to damping, much as
the fluctuations discussed previously.  We thus expect that their inclusion
generalizes the semiclassical result to $t \exp(-\Gamma t)\to K_1(t)\exp(-\Gamma
t)$.     

\subsubsection{Time scales $\sim D^L$ }

The analysis of non-perturbative effects at late scales requires a 
different strategy, which we here outline with additional details provided in
Appendix \ref{app:BeyondSemiclassics}.  

\paragraph*{Semiclassical exactness:} To prepare the
analysis of the survivor modes, we notice that in the non-interacting theory all
nonlinear contributions to the non-interacting actions \eqref{eq:S0DiscreteTime}
cancel out. This mechanism is known as `semiclassical exactness'. In
Appendix \ref{app:BeyondSemiclassics} we discuss how semiclassical exactness
extends to the interacting theory. The result of this discussion is that the sum
over quadratic actions Eq.~\eqref{eq:S0DiscreteTimeQuadratic} with the quadratic
reduction of the functional expectation value \eqref{eq:FormFactorMultiSiteB}
describes the expansion of the form factor to all order in perturbation theory
in $D^{-1}$. 

\paragraph*{Interactions:} The semiclassical exactness principle does not extend
to the expansion of the interaction vertex in $B$-fluctuations. We proceed by
splitting the interaction as $S_\textrm{int}=S_\textrm{int}^{(4)}+\delta
S_{\mathrm{int}}$ into a leading, quartic contribution plus rest. The quartic term
is then processed via the strategy outlined in Section~\ref{sec:Interactions}.
At an intermediate stage, we obtain a theory in which the functional expectation
value Eq.~\eqref{eq:FormFactorInHS} as well as non-synchronized contributions to
$\delta S_{\mathrm{int} }$ are weighed by phase factors containing the interaction
field $\Theta$. Since integration over these fields produces damping at scales
$\sim \Gamma^{-1}$, we ignore all contributions to the path integral carrying
$\Theta$-dependence.

\paragraph*{Late time effective theory:} At this point, the theory is reduced to
an expectation value
\begin{align}
    \label{eq:FormFactorBvsX}
    K(t)\to &D^{2L} \sum_{mn}\prod_j \left \langle (B_{tn} B_{n1}^\dagger B^\dagger_{1m} B_{mt})_j \right \rangle_B= \nonumber \\
   &=D^{2L}  \left\langle (X X^\dagger)_{t1} (X^\dagger X)_{1t}  \right\rangle_B,
\end{align} 
involving the globally synchronized `Hadamard mode'
Eq.~\eqref{eq:XModeDefinition}. It is now straightforward to establish the
latter as an effective integration variable, replacing the integration over
mulitple $B$-fields. As a result of a quick calculation detailed in Appendix
 \ref{app:IntegrationHadamardMode} we obtain 
\begin{align}
    \label{eq:FormFactorXQuadratic}
    K(t)\to D^{2L} \int DX \,  e^{-S_0[X] }  (X X^\dagger)_{t1} (X^\dagger X)_{1t},  
\end{align} 
with (cf. Eq.~\eqref{eq:S0DiscreteTimeQuadratic})
\begin{align}
    \label{eq:QuadraticXAction}
    S^{(2)}_0[X]=D^L\, \tr(X^\dagger d X),
\end{align}
i.e. the semiclassical  matrix action governing an ergodic quantum system with
spectral density $D^L$.
\paragraph*{Causal symmetry:}
This is as far as (infinite order) perturbation  theory around the
standard saddle point will get us. To complete the construction we need to
invoke  a symmetry argument. As discussed in Section~\ref{sec:Symmetries}, the
late time ergodic phase of our system is a symmetry broken phase in which
invariance under a large $\mathrm{U}(2t)$ is spontaneously broken to
$\mathrm{U}(t)\times \mathrm{U}(t)$, 
by the time translation
operator $d$ in the action. The Goldstone mode generator of this system-wide
global symmetry are precisely the matrices $X,X^\dagger $, combined into a
rotation matrix  (cf. Eq.~\eqref{eq:RationalParameterization}) 
\begin{align*}
   U=\begin{pmatrix}
    \mathds{1}&X\\ -X^\dagger & \mathds{1}
   \end{pmatrix}. 
\end{align*}  
The minimal theory extending Eq.~\eqref{eq:QuadraticXAction} to one
incorporating the symmetry principle is defined by
\begin{align}
    \label{eq:FullXAction}
    S_0[X]=\frac{D^L}{2}\, \tr(dU \tau_3 d U^{-1}),
\end{align}
i.e. the universal ergodic theory (cf. Eq.~\eqref{eq:SigmaModelInvariant}) of a
chaotic system with $D^L$ levels. This theory retracts to
Eq.~\eqref{eq:QuadraticXAction} to all orders in perturbation theory around the
standard saddle $\tau_3$, but also includes the AA saddle point structure
required to describe the plateau of the form factor at the many body Heisenberg
time $D^L$.

\section{Thermalization and symmetries}
\label{sec:Symmetries}

The  previous analysis demonstrated exponential relaxation towards an
 ergodic state on a time scale determined by the rate of
local correlations. However, the situation changes when  unitary
symmetries come into play~\cite{SpectralStatisticsManyBodyConservedCharge2019}.
 In this case, the continuity equations expressing the conservation of local
 charges lead to locally diffusive
dynamics, which takes a minimum time $\sim L^{2}$ to reach relaxation. Absent
inter-dot particle transport,  examples of such
 conserved quantities include  energy (in continuous time models) and spin. In
 the following, we consider the slightly  more  general symmetry under the action of a
continuous unitary transformation group, $G$.  We  begin by  exploring the consequences of such
symmetries within the framework of the discrete time model considered
previously. The important case of energy conservation (where $G$=group of time
translations) is addressed in section \ref{sec:HamiltonianSystems}.    

Consider a Lie group with Lie algebra generators $\{T^\alpha\}$, and
structure constants defined by $[T^\alpha,T^\beta]=f^{\alpha \beta \gamma}T^\gamma$. 
We wish to describe evolution  allowing for local exchange of the symmetry's conserved charges, 
at conserved total charge. With the local symmetry generators
defined as,
\begin{align}
    \label{eq:TGeneratorTensorProduct}
    T^\alpha_j \equiv \mathds{1}\otimes...  \otimes T^\alpha\otimes\dots,
\mathds{1}
\end{align}
acting on
qudit $j$, this condition is satisfied by the nearest neighbor exchange
operators $T^\alpha_j\otimes T_k^\alpha$, where summation over $\alpha$ is implied. These
operators do not commute with the locally represented symmetry, but do commute
with the total generators, 
\begin{align*}
    T^\alpha\equiv \sum_j T^\alpha_j,
\end{align*}
as a straightforward consequence of the antisymmetry of structure constants.
Assuming that the symmetry is acting on the single qudit level in a
$K$-dimensional representation, we tensor our fields $\psi_\mu\to
\{\psi_\mu^u\}$, $u=1,\dots,K$ and define the single qudit path integral
representation of the generators as $\bar \psi^u_{j,\mu}
T^\alpha_{uv}\psi_{j,\mu}^v$. 

We now declare that the interaction Hamiltonians acting as $\exp(i \delta
H_\textrm{int}) $ during the two-qudit evolution stretches of our network
dynamics include a symmetry exchange contribution as
\begin{align}
    \label{eq:HSymmetry}
    H_\textrm{s}=  \frac{\gamma}{2}\sum_{\langle j,k \rangle}(\bar \psi_j 
    T^\alpha\psi_{j})( \bar \psi_k
    T^\alpha\psi_k).
\end{align}
When included into our discrete time coherent state action
Eq.~\eqref{eq:SMicroscopic}, this operator appears as
\begin{align*}
    &S_\textrm{s}[\psi]=-\frac{i\Gamma_\textrm{s} }{2}
    \sum_{n,a} a\sum_{\langle j,k \rangle} 
    (\bar \psi^a_{jn-a} T^\alpha \psi^a_{jna}) (\bar \psi^a_{kn-a} T^\alpha \psi^a_{kna}),
\end{align*}
with $\Gamma_\textrm{s}\equiv \gamma_s \delta$.
Note that there is no randomness involved here, so we do have a contribution at
first order in the time increment $\delta$. 

Following our previous protocol, we next implement the locking
\eqref{eq:PsiGLockingDiscrete}. This is done by reordering the two factors in
the action as $(\bar \psi^a_{j-a} T^\alpha \psi^a_{ja})=-\tr(\psi^a_{ja}
\bar{\psi}^a_{j-a}T^\alpha)$, where the trace is over the $u$-representation space indices
of the symmetry. This  structure requires an upgrade $G\to \{G^{uv}\}$ of the
quasiclassical Green function to a matrix $u$-space, the intuition being
that $G^{uv}$ describes interfering trajectories carrying symmetry labels $u$
and $v$, respectively. The vertex in $G$-representation then reads
\begin{align}
    \label{eq:SSymmetriesG}
    S_s[G]=&-\frac{i\Gamma_\textrm{s} D^2}{2} \sum_{n,a} 
    a\sum_{\langle j,k \rangle} \tr(G^{aa}_{j,n-a,na} T^\alpha) \tr(G^{aa}_{k,na,n-a} T^\alpha).
\end{align}
We process this perturbation similarly to the interaction in Section
\ref{sec:Interactions}: an algorithm consisting of the expansion to leading
(quartic) order in $B$-generators, decoupling by a Hubbard-Stratonovich field,
and removal of the latter from the action via a --- now non-abelian --- gauge
transformation involving the $T^\alpha$-generators. Since this procedure will not introduce
conceptually new elements to our discussion, its detailed execution is relegated
to Appendix \ref{app:SymmetryExchange}. As a result, we obtain
the expression 
\begin{align}
    \label{eq:SFFFormFactorSymmetries}
    K(t)=t\,\left|\tr\left( e^{ it\Gamma_\textrm{s}\sum_{\langle j,k \rangle}T^\alpha_j T^\alpha_k} \right)\right|^2,
\end{align}
which structurally resembles Eq.~\eqref{eq:FormFactorDiscreteTime},
i.e. the semiclassical SFF as a trace over a configuration space. Presently, we
sum over  local spin configurations, with a statistical
weight given by $i t\times$ the spin exchange operator. 

Physically, Eq.~\eqref{eq:SFFFormFactorSymmetries} multiplies the `spin-singlet'
SFF with  the squared
time evolution operator of  an $\mathrm{SU}(N)$-Heisenberg model.  
This particular result is the consequence of a symmetry of our model,
namely the commutativity of the Hamiltonian Eq.~\eqref{eq:HSymmetry}, with the
Floquet operator describing the unperturbed dynamics. This means that the model
actually is trivial in the sense that the factorization of the time evolution operators visible in Eq.
\eqref{eq:SFFFormFactorSymmetries} is an exact feature. It is relatively
straightforward to extend our solution strategy to cases where this symmetry is
absent, e.g., due to position dependent, or random couplings.
However,  the general conclusions below extend to models generalized in this
way, and we do not discuss them explicitly.

A similar result has been obtained for $K=2$ in
Ref.~\cite{SpectralStatisticsManyBodyConservedCharge2019}. In either case, the
slow dynamics of local spin exchange is the bottleneck for relaxation  towards the ergodic
phase. Generically, the relaxation time scale, commonly called the Thouless
time, $t_\textrm{Th}$, will scale as $t_\textrm{Th}\sim L^2$, reflecting the
energy dispersion of the lowest lying `magnon' excitations above the ground
state of the model. However, there can be exceptions to this rule. For example,
anti-ferromagnetic coupling realized in a one-dimensional system tuned to have
interaction couplings alternating as $\Gamma_s\sim (-)^l$, would lead to a
shorter relaxation time $t_\textrm{Th}\sim L$ reflecting the linear dispersion
of antiferromagnetic couplings. Another scenario, realized for $K=3$ chains with
antiferromagnetic coupling, is the presence of an excitation gap in the system.
In this case, we should expect relaxation towards the ground state at a time
scale independent of system size. 

Summarizing, the presence of symmetries and their charges satisfying local
continuity equations generically slows the relaxation of complex quantum
systems, as pointed out in
Ref.~\cite{SpectralStatisticsManyBodyConservedCharge2019}. In the next section,
we investigate how this principle works in the case of time translational symmetry.

\section{Hamiltonian systems}
\label{sec:HamiltonianSystems}

\begin{figure}[h]
    \centering
    \includegraphics[width=0.5\linewidth]{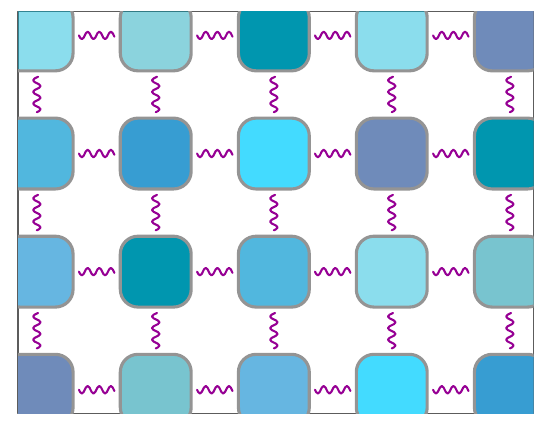}
    \caption{An array of random quantum dots coupled by pair interaction. 
    The color shading indicates  statistical independence of the on-site 
    Hamiltonians.}
    \label{fig13}
\end{figure}

In this section, we extend our approach to Hamiltonian systems. Instead of the previously discussed 
array of unitary circuits we consider a system of $L$ `quantum dots' with
$D$-dimensional Hilbert space governed by independently drawn Hamiltonians $h_l$ with
variance 
\begin{align*}
    \langle |h_{\mu \rho}|^2 \rangle = \frac{\lambda^2}{2D},
\end{align*}
coupled by the two-body interaction Eq.~\eqref{eq:InteractionVariance}, cf. Fig.~\ref{fig13}. 
 Most of the changes relative to the previously considered
system amount to taking straightforward continuum time limits. However,  
the  non-uniform spectral density of Hamiltonian systems also leads to a
few complications. They  relate to structures at the scale of the energy
cutoff $\lambda$ of individual sites, over which the theory has parametric 
but no numerical control. Where possible, we
will fix numerical factors by consistency arguments, but otherwise leave them unspecified.

\subsection{Luttinger-Ward functional}

We start by adapting elements of the
previous discrete time formalism to the present setting.  For example, with
$U=\exp(iH \delta)$, and $\delta$ an infinitesimal
time step, the continuum version of the action in
Eq.~\eqref{eq:SingleQuditPartitionSum} reads as
$S[\psi]=\int dt \,\bar \psi(\partial_t - i H)\psi$ and the interacting theory
is described 
by (cf. Eq.~\eqref{eq:SMicroscopic}) by
\begin{align*}
    S[\psi]=\sum_a a\int_t &\bigg( \sum_j \,\bar \psi_j^a(\partial_t - i h_j)\psi_j^a \\
    &- i
     \sum_{\langle j,k \rangle} \bar \psi^a_j \bar \psi^a_k H_{\textrm{int}} \psi^a_k \psi^a_j\bigg), 
\end{align*}
with continuum fields $\psi=\{\psi_{j,t}^a\}$, i.e. by an  orthodox coherent
state field integral action. The subsequent averaging over the ensemble
generates quartic ($h$) and eighth order ($H_\textrm{int}$) fermion terms, to
which we apply the replacement rule Eq.~\eqref{eq:QuasiaClassicalGreenFunction},
and then integrate over Grassmann fields. As a result, we obtain the continuum
version of the $G\Sigma$-action,
\begin{align}
    \label{eq:ContinuumGSigma}
    S[G,\Sigma]&=-D \sum_j \left(\tr\ln\left( i \partial_t - \Sigma\right) +
      \tr\left(\Sigma G- \frac{\lambda^2}{2} G^2 \right)\right)_j\nonumber\\
    &+\frac{\Lambda^2 D^2}{2}\sum_{\langle j,k \rangle}\tr\left( (G_j\odot G_k)\tau_3 (G_j\odot G_k)\tau_3 \right),
\end{align}
where the second line defines the interaction action $S_\textrm{int}[G,\Sigma]$
and the  traces now include summation over the causal index, and integration
over time, i.e. $\tr(X)=\sum_a \int_t X^{aa}_{tt}$.

\subsection{Stationary phase analysis} 

Varying Eq.~\eqref{eq:ContinuumGSigma}, we obtain the equations 
\begin{align*}
    0
    &=
    \delta_{\Sigma_j} S[\bar G,\bar \Sigma]
    =
    \bar G_j-\frac{1}{i\partial_t - \bar \Sigma_j},
    \\
0
&=
\delta_{G_j}S[\bar G,\bar \Sigma]
=
D \lambda^2 \bar G_j - D \bar \Sigma_j +\delta_{G_j}S_{\textrm{int}}[\bar G].
\end{align*}
If it were not for the  interaction $S_\textrm{int}$, these would define a set of
self-consistency equations for a system of decoupled disordered quantum dots, 
identifying the on-site self energy $\Sigma_j = \lambda^2 G_j$ 
with the Green function. Before turning to the role played by interactions, we
solve this equation, temporarily omitting the site index $j$. Turning to
Fourier space (cf. the conventions stated above
Eq.~\eqref{eq:PairCorrelationFunction}), $i\partial_t \to \epsilon$,  the
equation assumes the form of a self-consistent Born equation
\begin{align}
    \label{eq:SCBAEquation}
    \bar G = \frac{1}{\epsilon-\lambda^2 \bar G}.
\end{align}
We focus on energies $|\epsilon|<\lambda$, for which the equation is
approximately solved by
$\bar G^a \simeq -i a \lambda^{-1} $, where causality
determines the retarded and advanced choice of sign $a$. 
Putting $\epsilon$ back
in we obtain $\bar G^a \simeq \frac{1}{\epsilon + i a\lambda}$, and  Fourier
transformation  back to the time domain leads to 
\begin{align}
    \label{eq:GMeanFieldContinuousTime}
    \bar G^a_{tu}= -ia \Theta (a(t-u))e^{-\lambda |t-u|}.
\end{align} 

We next demonstrate that Eq.~\eqref{eq:GMeanFieldContinuousTime} defines a
solution of the stationary phase equations, including in the presence of
interactions. To this end, consider $S_\textrm{int}[\bar G]\equiv
\frac{\Lambda^2D^2}{2}\sum_{\langle j,k \rangle}\sum_a  \int_{tu}
((\bar{G}^a_j\odot \bar G^a_k)_{tu} (\bar G^a_j\odot \bar G^a_k) )_{ut}$, where we
noted that the mean field solution is diagonal in causal $a$-space. The time
integral extends over products of Green functions of identical causality
evaluated at positive and negative time differences, implying the vanishing of
both, this expression, and of the derivative $\delta_{\bar G}S_\textrm{int}$.
Physically, this finding reflects a  vanishing contribution of the interaction
to the self energy. The reason is that an interaction contribution to the single
particle propagator would necessitate the creation of a ``particle-hole
excitation'' in a neighboring qudit, which is not an option in our theory with
single particle Hilbert space occupation.   

\subsection{Continuum action} 

As discussed in section~\ref{sec:Essentials}, the presence of
Goldstone modes in our system is a consequence of the symmetry breaking between
$a=\pm1$ sectors on the mean field level. Formally, the  Goldstone mode
manifold is generated by continuous transformations of the mean
field $\bar G \to T \bar G  T^{-1} $, where $T=\{T^{ab}_{tu}\}$ are
matrices in causal and time domain. In order to avoid ``empty integrations'', it
is customary to represent these modes as $T=\mathds{1}+W + \dots$, where the
generator matrices $W$ are chosen to anti-commute with the mean field $\bar G$.

Differently from the discrete model, the mean field $\bar G_{t-u}$ carries
explicit time dependence, making  the  anti-commutativity condition difficult to realize exactly. At the
same time, $\bar G_{t-u}\approx -i\tau_3 \delta_{t-u}$ decays rapidly as a
function of distance, and fields $W_{tu}$ with time variation on scales
$>\lambda^{-1}$ are oblivious to the finite range of $\bar G$. Imposing this
 slowness condition, we continue to work with the representation
Eq.~\eqref{eq:RationalParameterization}, where the $B$-dependent generator
anti-commutes with $\bar G$ in the $\delta$-function approximation. At the same
time, we must not forget about the time-dependence of $\bar G$ entirely; the
causality condition $\bar G^a_{t-u}\propto \Theta (a(t-u))$ remains an
essential element in the derivation of the theory.

\paragraph*{Single site action and form factor:} The computation
of the continuous time effective action, detailed in Appendix
\ref{app:ContinuousTime}, parallels  our previous discrete time
construction and yields
\begin{align}
    \label{eq:SNoninteracingContinuous}
 S_0[T]=- \frac{D}{\lambda} \sum_j \int_{tu}\tr\left( \tau_3 T_{tu}^{-1} (\partial_{u}+\partial_t)T_{ut}\right)_j,
\end{align}
for the non-interacting action. Upon insertion of the representation
Eq.~\eqref{eq:RationalParameterization} this becomes the continuous time version of
Eq.~\eqref{eq:S0DiscreteTime}. (The role of the discrete derivative $d$ in the
center coordinate is now taken by the sum of two time derivatives.) For our
present purposes, it will be sufficient to consider the quadratic expansion in
$B$'s, which is readily obtained as
\begin{align}
    \label{eq:S0Continuum}
    S_0^{(2)}[B]&=\frac{2D}{\lambda}\sum_j\int_{t u}B^\dagger_{t u}(\partial_{t}+\partial_{u})B_{u t}=\nonumber\\
    &=\frac{2D}{\lambda}\sum_j\int_{t\Delta t}B^\dagger_{t\Delta t}\partial_t B_{t\Delta t},
\end{align}
and now assumes the role of Eq.~\eqref{eq:S0DiscreteTimeQuadratic}. In the
second line, we switched to a center ($t=(t + u)/2$) and difference time
($\Delta t= t - u$) representation $B_{tu}\to B_{t \Delta t}$,
$B^\dagger_{u t}\to B^\dagger_{t,\Delta t}$, which will be convenient
throughout. The continuum time version of the propagator
Eq.~\eqref{eq:BPropagatorDiscrete} is given by Eq.~\eqref{eq:BPropagator}, and
that  of the SFF~\eqref{eq:FormFactorMultiSiteB} by
\begin{align}
    \label{eq:FormFactorContinuous}
    K(t)\sim \left( \frac{D}{\lambda} \right)^{2L} \prod_k \left \langle (B B^\dagger)_{kt0} (B^\dagger B)_{k0t} \right \rangle_B, 
\end{align}
where the factor $\lambda^{-2L}$, formally introduced by the replacement
rule~\eqref{eq:GvsQ}, balances the dimension of field matrix elements,
$[(BB^\dagger)_{tu}]=\,\textrm{(energy)}$.

\paragraph*{Interaction vertex:} Expanding  the interaction $S_\textrm{int}[G]$ to quartic order in
fluctuation generators $B$ we obtain the continuous time variant of Eq.~\eqref{eq:SintLowestOrder} (cf. Appendix \ref{app:ContinuousTime})
\begin{align}
    \label{eq:SIntQuarticContinuum}
    &S^{(4)}_\textrm{int}[B]= \frac{2\Gamma D^2}{\lambda}   \sum_{\langle j,k \rangle}
     \int B_{jt_+ t_-}B_{ku_+ u_-} B^\dagger_{jt_- t_+}B^\dagger_{ku_- u_+}\nonumber\\
  &\qquad \times \left(\lambda^{-1}  \delta_{t_+,u_+}+\lambda^{-1}\delta_{t_-,u_-}- 8\lambda^{-2} \delta_{t_+,u_+}\delta_{t_-,u_-} \right),
\end{align}
where $\Gamma = 4 \Lambda^2 /\lambda$ has dimension energy and $\int$ is shorthand
notation for the integration over all four  time variables. Here,  the
$\delta$-functions are to be interpreted as smeared over scales $\lambda^{-1}$.
On this basis, we observe the synchronization principle at work again: for field
configurations $B_{jt_+ t_-}B_{lt_+ t_-}$ with identical time arguments (always
with $\lambda^{-1}$ uncertainty) the two terms cancel out, stabilizing a
global ergodic mode. 

However, differently from the discrete model, uniform time translations  now
realize  a continuous symmetry, which leads to qualitative effects in the
late time dynamics.
Comparing to our earlier discussion in Section~\ref{sec:Symmetries}, the role of
the symmetry generators $T^a$ is now taken by the time translation operator,
$\partial_t$, energy being the  globally conserved charge. 

These  parallels  are best exposed in a hybrid energy-time
representation defined by the Wigner transform of generators, 
\begin{align*}
    &B_{t\epsilon}\equiv \int_{\Delta t}e^{i \Delta t \epsilon}B_{t+\frac{\Delta t}{2}t-\frac{\Delta t}{2}},\\
    &B_{t+\frac{\Delta t}{2}t-\frac{\Delta t}{2}}=\int_\epsilon \, e^{-i \epsilon \Delta t}B_{t\epsilon}.
\end{align*}
It is straightforward to verify that in the Wigner language the interaction assumes the form
\begin{align}
    \label{eq:SinteractionWigner}
    S_\textrm{int}^{(4)}[B]&=4\Gamma D^2\sum_{\langle j,k \rangle} \int
    \Big( B_{jt\bar \epsilon + \frac{\Delta \epsilon}{2}}B^\dagger_{jt\bar \epsilon - \frac{\Delta \epsilon}{2}} 
    B_{kt\bar \epsilon + \frac{\Delta \epsilon'}{2}}B^\dagger_{kt\bar \epsilon - \frac{\Delta \epsilon'}{2}}  \Big)\nonumber\\
   & \qquad\qquad\left(\delta_{\Delta \epsilon - \Delta \epsilon '}- 8 \lambda^{-1}  \right),
\end{align}
where we assumed negligible dependence of the fields on the center time $t$ on
microscopic time scales $\sim \lambda^{-1}$. 

\begin{figure}[h]
    \centering
    \includegraphics[width=0.75\linewidth]{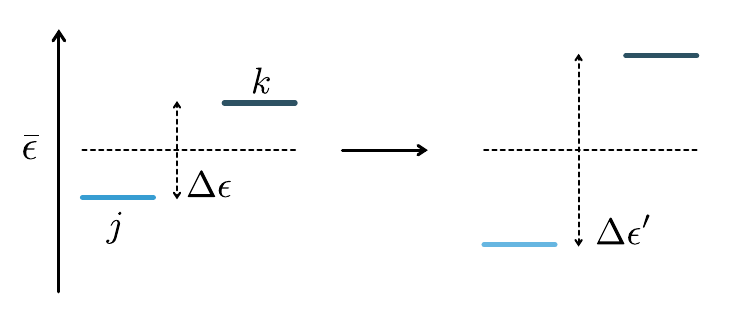}
    \caption{Energy exchange between neighboring dots $j,l$ as described by the
    interaction vertex Eq.~\eqref{eq:SinteractionWigner}.
  }
    \label{fig14}
\end{figure}

Thinking of the fields $B_{jt\epsilon}$ as  measures for the probability for
site $j$ to be in a state with energy $\epsilon$ at time $t$, this vertex
describes  stochastic dynamics in which neighboring dots exchange energy from
configurations with local energy difference $\Delta \epsilon$ to $\Delta
\epsilon'$, at conserved center energy $\bar \epsilon$, cf. Fig.~\ref{fig14}. Repeated of these energy
transfer processes describe the diffusive relaxation of energy in the system.

\subsection{Energy diffusion} 

In view of the structures outlined above, we expect the emergence of an
effective action playing a role analogous to the previously discussed
$\textrm{SU}(N)$-Heisenberg action. Time translational invariance is microscopically realized as
$\psi_t \to (e^{T^a}\psi)_t=\psi_{t+a}$,  the generators $T^a=a \partial_t$
assuming the role of the previously considered $\textrm{SU}(N)$-generators.

To identify the action cost of weakly position-dependent symmetry operations,
$a\to a_j$, we follow a continuum time version of the
discrete time protocol outlined in Section~\ref{sec:Interactions}.  Referring to
Appendix~\ref{app:InteractionContinuousTime} for its detailed execution, it
yields the
form factor as 
\begin{align}
    \label{eq:SFFTimeFunctional}
    K(t)=  &\lambda^L \int D\tau \, e^{-S[\tau]}, \nonumber \\ 
    & S[\tau]=-4 \Gamma  t  \sum_{\langle j,l
\rangle} F(\tau_j-\tau_l),
\end{align}
where the integration $\int D \tau=\prod \int d\tau_j$ extends over the relative
time coordinates of the individual dots. The weight function, $F=1-f$, where $f(\tau)$ is a  Gaussian
implicitly defined by $f(0)=1$, and $f(\tau\gg \lambda^{-1})=0$. 

Eq.~\eqref{eq:SFFTimeFunctional} is the continuous time version of
Eq.~\eqref{eq:FormFactorDiscreteTime}. The only difference is that the cost
function for dots out of temporal synchronization is now continuous, instead of
step-function like. In the short time limit, $\Gamma t\to 0$, this leaves us
with $L$ independent time integrations, and the product form factor $K(t)\sim
(\lambda t)^L$. For $\Gamma t\gg 1$, the action enforces near-synchronization
to time configurations $\tau_l = \tau+ \lambda^{-1} a_l$, where the
dimensionless time-mismatch parameters $a_l$ are defined to have zero mean,   $\sum_l a_l=0$.
 To describe the manifestation of time translational invariance in the SFF, we            
note that  for $\Gamma t \gg 1$, the
synchronization mismatches $|a_j-a_l|=\mathcal{O}(1)$ must be
small.  A second order Taylor expansion
$F(a)\sim  a^2$, then gets us to the approximation
\begin{align}
    \label{eq:ContinuousTimeAAction}
    S[a]\sim t \Gamma \sum_{\langle j,l \rangle} (a_j-a_l)^2
    \sim t \Gamma  \int_0^L dx \, (\partial_x a)^2,
\end{align}
where  the  continuum limit introducing  a dimensionless length
parameter is meaningful  in the  case where the number of
sites is large. 

Eq.~\eqref{eq:ContinuousTimeAAction}  describes the relaxation of the system at large time scales.  To
estimate the contribution of its 
lowest lying excitations to the SFF, we recall that the integration $\int Da$ extends
over fluctuations excluding the constant mode $\sum_l a_l=0$. We
implement this condition by passing to an integration over momentum modes, 
\begin{align*}
    a_q = \frac{1}{L^{1/2}}\sum_j e^{-iqj}a_j,\qquad a_j = \frac{1}{L^{1/2}}\sum_q e^{iq j }a_q,
\end{align*} 
where the normalization prefactor is defined in such a way that the passage
$\int Da=\int \prod_j a_j\to \int \prod_q da_q$ has unit
determinant, and the product extends over all $q$-values
$q=(2\pi/L)(1,\dots,L-1)$, excluding the constant mode $q=0$. In the
$q$-representation, the action assumes the form 
\begin{align}
    \label{eq:TimeTranslationModeAction}
    S[a]\sim t \Gamma  \sum_{q} \sin^2 \left( \frac{q}{2} \right)  |a_q|^2,
\end{align}
and for the  modes $q=\pm 2\pi/L \,\textrm{mod} \,2\pi$ with lowest action. A 
linearization of the $\sin $ yields
\begin{align*}
    S[a_{2\pi/L}]\sim \Gamma\frac{t}{L^2}|a_{2\pi/L}|^2,
\end{align*} 
demonstrating the  scaling $\sim t/L^2$ characteristic for a diffusive
relaxation process. To describe its influence on the form factor, we integrate
over $q$-modes, to obtain Eq.~\eqref{eq:ContinuumProductFormula},
where the cutoff imposed by the $\textrm{max}$-function
reflects the lack of resolution of the lowest time scales $a_q\sim
1$ in our theory: For large values of $t$, the exponentials $\exp(-\Gamma t\sin^2(q)|a_q|^2)$ appear to suppress the integration to
windows of width $\Delta a^2 \sim (\sin^2(q) t \Gamma)^{-1} \lesssim
1$. However, these lie below the time-uncertainty implied by the
presence of microscopic fine structures, notably the mean field Green functions,
of range $\lambda$. In fact, one can show by a somewhat elaborate diagrammatic
analysis not reproduced here, that fluctuations of $\mathcal{O}(\lambda^{-1})$
in the $a$-variables of the pre-exponential sources do not significantly
affect the value of the form factor. We therefore implicitly assume  a small tolerance
window in the definition of the weight function in Eq.~\eqref{eq:SFFTimeFunctional}.

As discussed in Section~\ref{sec:SummaryOfResults},
 the approach to the ergodic form
 factor described by  Eq.~\eqref{eq:ContinuumProductFormula} is softer than 
 the exponential relaxation in the model without symmetries. Presently, the
 long-lived modes responsible for the slow decay are the weakly inhomogeneous
 time-translation modes, as witnessed by the action Eq.~\eqref{eq:TimeTranslationModeAction}.

\section{Summary and discussion}
\label{sec:Summary}

In this paper, we introduced a microscopic path integral describing the
evolution of complex  systems whose irreversible yet unitary dynamics eventually
leads to a maximum entropy `thermal' state. Our approach started
with the identification of  $D$-dimensional subsystems defined to be chaotic and
quasi-instantly relaxing into an ergodic state if kept in isolation. We
described individual of these systems in terms of quasiclassical Green
functions, degrees of freedom tracking the times at which
pair amplitudes  propagate in their ergodic
backgrounds. Interactions between the subunits introduced correlations, causing the locking 
of previously independent  amplitudes to a single effective many-body state
propagating in the multi-system tensor product space. For 
interaction rates $\Gamma$, this initial stage of the dynamics is completed
after a short (sub-extensive) timescale 
$\sim \Gamma^{-1}f(L,D)$, where $f$ is logarithmic in 
system size and Hilbert space dimensions. We pointed out that for interaction
rates $\Gamma\lesssim D^{-1}$, the resolution of this crossover dynamics required
resolving fluctuations non-perturbative in $D$, and hence outside the
range of semiclassical approaches.  

For systems with local symmetries, the subsequent relaxation dynamics of 
conserved charges via local exchange processes reaches an ergodic
state after time scales $\sim L^2$ characteristic for diffusive dynamics. We
exemplified this mechanism on two case studies, circuit arrays with
$\mathrm{U}(N)$ rotation symmetry, and Hamiltonian dynamics with time
translational symmetry.

The path integral in this paper is defined over fields in Fock space, unlike
conventional formulations in $d+1$-dimensional space-time. In this context, our analysis revealed
a conceptual challenge without a current resolution: causality necessitates two
distinct symmetry-breaking mechanisms—one at the subsystem level, the other at
the global system level. These symmetries break and are restored on different
time scales, set by the Hilbert space dimensions $D$ and $D^L$, respectively.
Two alternative path integral formulations using many-body coherent
states—reflecting a choice between first- and second-quantized dynamics—capture
one of these mechanisms, but not both (see
Section~\ref{sec:SymmetriesVsInteractions}). This highlights the challenge of
describing strongly entangled states using separable coherent product states;
resolving it could open the door to a more effective analytical framework for
complex quantum dynamics.

However, even in its current form, the path integral framework developed here enables an
efficient, microscopically resolved description of quantum relaxation in a broad
class of systems with integrability-breaking randomness.\footnote{ (Describing
the evolution of systems with intrinsic integrability breaking can be harder but
does not pose a conceptual problem. As an example, we mention rotor models,
where the path integral of individual kicked rotors is under excellent
control~\cite{ianTheoryLocalizationResonance2010a}. Coupled rotor models can be
approached by a relatively straightforward adaption of the construction applied
in Section~\ref{sec:CircuitNetworks}).} We are confident that this framework
will evolve into a versatile toolbox for tackling problems in many-body quantum
chaos that lie beyond the reach of numerical or semiclassical perturbative
methods.

\paragraph*{Acknowledgments:} We acknowledge 
financial support by Brazilian agencies CNPq and FAPERJ, and
partial support from the Deutsche
Forschungsgemeinschaft (DFG) under Germany’s Excellence
Strategy Cluster of Excellence Matter and Light for Quantum
Computing (ML4Q) EXC 2004/1 390534769 and within the
CRC network TR 183 (project grant 277101999) as part of
projects A03. 
K.W.K. acknowledges financial support from the Basic Science Research Program through the National Research Foundation of Korea (NRF) funded by the Ministry of Education (no. RS-2025-00521598) and the Korean Government (MSIT) (no. 2020R1A5A1016518). 
 Data and materials availability: Processed data and python script used to generate
Fig. 2,3,4 are available in Zenodo with identifier 10.5281/zenodo.15782178.

\appendix

\section{Grassmann path integral}
\label{app:GrassmannPathIntegral}

We here discuss the construction of the  path integrals discussed in the main
text, beginning with the case of single qudit evolution, Section
~\ref{sec:SingleQudit}.

\subsection{Single qudit path integral}

For  a single Grassmann variable, $\psi$,~\cite{Altland2023}  consider the definition of the
 integral
\begin{align*}
    \int d\psi \, \psi^{m}= \delta_{m,1}.
\end{align*}
It implies the vector integral generalization
\begin{align}
    \label{eq:GrassmannKronecker}
    \delta_{\mu \rho}=\int D\psi\, e^{-\bar \psi \psi}\psi_{\mu} \bar \psi_{\rho}
\end{align}
where $\psi= \{ \psi_{\mu} \}  $ now is a $D$-component Grassmann vector, 
$D\psi=\prod_\mu d\bar \psi_\mu d\psi_\mu$ and $\bar \psi \psi 
=\sum_\mu \bar{\psi}_\mu \psi_\mu$. 

Using these equations, we represent matrix elements of the time evolution
operator with redundantly introduced Kronecker-$\delta$'s:
\begin{align*}
    &\braket{\rho|U^t|\mu}=\delta_{\mu_{t+1}\rho} \delta_{\mu_{t+1} \rho_t} U_{\rho_t \mu_t}\delta_{\mu_t \rho_{t-1}}
    \dots U_{\rho_1 \mu_1}\delta_{\mu_1 \rho_0} \delta_{\rho_0 \mu} =\\
   & \quad\delta_{\mu_{t+1}\rho} \int D\psi e^{-\bar \psi \psi} \psi_{t \mu_{t+1}}\bar \psi_{t \rho_t}U_{\rho_t \mu_t} \psi_{(t-1)\mu_t}\bar \psi_{(t-1)\rho_{t-1}}\dots \\
   &\hspace{3cm} \dots U_{\rho_1 \mu_1}\psi_{0 \mu_1  }\bar \psi_{0 \rho_0}\delta_{\rho_0 \mu}=\\ 
   &\quad=\int D \psi\, e^{-\bar \psi \psi}  \psi_{t \rho}\, e^{ \sum_{n=1}^t \bar \psi_{n}U \psi_{n-1}}\,\bar\psi_{0 \mu}=\\ 
   &\quad =\int D \psi\, e^{-\bar \psi \psi+ \bar \psi U T_- \psi}   \psi_{t \rho}\bar\psi_{0 \mu}.
\end{align*} 
Here, the individual Kronecker tensors are resolved by Grassmann integrals over
$\{\psi_{n,\mu}\}$. Introducing an analogous representation for $\tr(U^{\dagger
t})$, we obtain the result Eq.~\eqref{eq:SingleQuditPartitionSum}.

\subsection{Qudit network}

The interaction contribution can be treated analogously, by insertion of the
 qudit generalization of Eq.~\eqref{eq:GrassmannKronecker},  
\begin{align}
    \label{eq:GrassmannKronecker2}
    \delta_{\mu \nu, \rho \sigma}=\int D\psi\, e^{-\bar \psi \psi} \psi_{2 \nu}\psi_{1\mu}  \bar \psi_{1\rho}  \bar \psi_{2 \sigma}.
\end{align}  
(Notice the `anti-normal ordering',  grouping creation and annihiliation
`operators' together.) If we now place such Kronecker-$\delta$ representations
at the interaction time slices $(m,-1)\to (m,1)$ on
the bonds of the tensor network defined by Fig.~\ref{fig10}, we encounter Grassmann contractions of the
type
\begin{widetext}
\begin{align*}
        &\bar \psi_{j1\rho}\bar \psi_{k1\sigma}
    V_{\rho\sigma,\mu\nu}\psi_{k-1\nu} \psi_{j-1\mu}\approx \bar \psi_{j1\rho}\bar \psi_{k1\sigma} 
  \left( \delta_{ \rho\mu}\delta_{\sigma\nu}+i  X_{\rho\sigma,\mu\nu} \right)\psi_{k-1\nu} \psi_{j-1\mu}
   \to e^{\bar \psi_{j1 }\psi_{j-1 }+ \bar \psi_{k1 }\psi_{k-1} + i  \bar \psi_{j1}\bar \psi_{k1}X\psi_{k-1} \psi_{j-1}} , 
\end{align*} 
\end{widetext}
where $X\equiv 
H + \frac{i }{2}H^2 $. Combining all these factors into a
single exponent, we arrive at the interaction action $S[V,\psi]$. 

\section{Discrete time interaction vertex}
\label{app:LuttingerWard}
In this Appendix, we provide some details on 
the interaction vertex of the discrete time theory.

\subsection{The matrix $\mathcal{G}(B)$}

The construction of the path integral starts with the integration over
$\psi$-variables, Gaussian after all nonlinear terms have been Lagrange
multiplier locked to $G$. Collecting the quadratic terms in
Eqs.~\eqref{eq:SMicroscopic}, \eqref{eq:BPsiAction},  we obtain the bilinear
form $\bar \psi \mathcal{G}^{-1}\psi$ with
\begin{align}
    \label{eq:CalGDefinition}
\mathcal{G}^{-1}(B)=
  \begin{pmatrix}
  1&& B_T &\cr
  -1&1&&\cr
  &&1&-1\cr
  &- B^\dagger&&1\cr
  \end{pmatrix}
\end{align}
  where the matrix structure refers to a representation of the integration
  variables as $\psi= ( \psi_{-1}^+, \psi_{1}^+, \psi_{-1}^-, \psi_{1}^-)^T$, and
  the site-indices $j$ are left implicit. The
  straightforward inversion of $\mathcal{G}^{-1}$, e.g. by series expansion
  around the unit-matrix diagonal, yields Eq.~\eqref{eq:CalGDefinition}. Here,
  the unit operators on the diagonal  
 
  Adding to this bilinear form the quadratic term from \eqref{eq:LagrangeMultiplier}, we obtain
  $\bar \psi (\mathcal{G}^{-1}(B)-i\Sigma \tau_3)\psi$. The integral over $\psi$
  then leads to the second tr ln in Eq.~\eqref{eq:GSigmaAction}.

\subsection{Interaction vertex}
\label{app:InteractionsHigherOrder}

In this Appendix, we discuss the expansion of the discrete time interaction
vertex in $B$-fluctuations. Our starting point is the configurations
Eq.~\eqref{eq:StationaryPhaseArray}. With Eq.~\eqref{eq:CalGDefinition}, and
\eqref{eq:QMatrixB}, a breakdown of the individual components we need in the
following is given by  
(discrete time indices omitted for brevity)
\begin{alignat}{4}
    \label{eq:GvsQDictionary}
    & \bar{G}^{++}_{-1,1}\to \frac{i}{2}(Q^{++}-1)&&= -\frac{iB_TB^\dagger}{1+B_TB^\dagger},\nonumber\\
    & \bar{G}^{--}_{1,-1}\to \frac{i}{2}(Q^{--}+1)&&= \frac{i B^\dagger B_T}{1+B^\dagger B},\nonumber\\ 
    & \bar{G}^{+-}_{-1,-1}\to \frac{i}{2} Q^{+-}&&=-\frac{iB_T}{1+B^\dagger B_T},\nonumber\\
   &  \bar{G}^{-+}_{1,1}\to \frac{i}{2}Q^{-+}&&=-\frac{i B^\dagger}{1+B_TB^\dagger}.
  \end{alignat}
We substitute these terms into the interaction Eq.~\eqref{eq:InteractionG},
temporarily dropping the $\langle j,k \rangle$ summation over neighboring sites,
and relabelling the order parameter fields as $Q_j\to Q, Q_k \to R$. Their
Goldstone mode generators will be denoted by $B_{Tj}\to B$, $B_{Tk}\to C$. In this
notation, the interaction vertex assumes the form 
\begin{widetext}
    \begin{alignat}{4}
        \label{eq:InteractionVertexExpanded}
        S_{\textrm{o}}[B,C]&=\frac{\Gamma D^2}{4}\tr\left(
        (Q^{++}-1)\odot(R^{++}-1)+ (+\leftrightarrow -)  \right)=\\\nonumber
        &=\Gamma D^2\sum_{n}\sum_{q,r=1}^\infty(-)^{q,r}\left(
        (BB^\dagger)_{nn}^{q}(CC^\dagger)_{nn}^{r}+(B^\dagger
        B)_{nn}^{q}(C^\dagger C)_{nn}^{r} \right),\\ \nonumber
        S_{\textrm{i}}[B,C]&=\frac{\Gamma D^2}{16}\tr\big(
        [(Q^{++}-1)\odot (R^{++}-1)][(Q^{++}-1)\odot (R^{++}-1)]- [Q^{+-}\odot
        R^{+-}][Q^{-+}\odot R^{-+}]+ (+\leftrightarrow -)  \big)=\\\nonumber
        &=D^2\sum_{n,m}\sum_{p,q,r,s=0}^\infty(-)^{p+q+r+s}\\\nonumber
        &\qquad \big((B B^\dagger)_{mn}^{p+1}(CC^\dagger)_{mn}^{q+1}(B
        B^\dagger)_{nm}^{r+1}(C C^\dagger)_{nm}^{s+1}+
        (B^\dagger B)_{mn}^{p+1}(C^\dagger C)_{mn}^{q+1}(B^\dagger B)_{nm}^{r+1}(
        C^\dagger C)_{nm}^{s+1}-\\ \nonumber
        &\qquad - (B (B^\dagger B)^p)_{mn} (C (C^\dagger C)^q)_{mn}(B^\dagger ( BB^\dagger )^r)_{nm} (C^\dagger ( CC^\dagger )^s)_{nm}-\\ \nonumber 
        &\qquad - (B^\dagger ( B B^\dagger B)^p)_{mn} (C^\dagger (C C^\dagger )^q)_{mn}( B ( B^\dagger B )^r)_{nm} (C ( C^\dagger C )^s)_{nm}\big).
    \end{alignat}   
\end{widetext}
Extracting the terms with $p=q=r=s=0$, and turning back to the original notation
    $B\to B_{Tj}$, $C\to B_{Tk}$ we obtain the fourth order vertex
    Eq.~\eqref{eq:SintLowestOrder} discussed in the main text. 
    For the discussion of terms beyond quartic perturbation theory, we refer to
Appendix \ref{app:InteractionBeyondQuartic}.

\section{Hubbard-Stratonovich decoupling of the interaction}
\label{app:HSInteraction}

In this Appendix, we discuss the decoupling of the in-interaction according to
the procedure outlined in section \ref{sec:Interactions}. To keep the notation
slim, we interchangeably use notation $B_{n_+n_-}\equiv B_n$ and
$B^\dagger_{n_-n_+}=B^\dagger_n$ throughout.   We start with a compactified
representation of the interaction Eq.\eqref{eq:SintLowestOrder}:
\begin{align}
    \label{eq:SIntCompactDiscrete}
        S_\textrm{int}^{(4)}[B]&= \Gamma D^2\sum_{\langle j,k \rangle,nm}(B_{T}B^\dagger)_{jm} 
        F_{mn} M_{jk} (B_{T}B^\dagger)_{kn},    
\end{align} 
where $(B_TB)_{jm} \equiv B_{Tjm} B^\dagger_{jm}= B_{Tjm_+ m_-}B^\dagger_{jm_-m_+}$,
\begin{align}
    \label{eq:MatrixMDef}
    M_{jk}=\begin{cases}
        1&(j,k) \,\,\, \textrm{neareast}\,\textrm{ neighbors},\\
        0&\textrm{else},
    \end{cases}
\end{align}
and 
\begin{align}
    \label{eq:FDefinitionDiscrete}
    F_{nm}\equiv \delta_{n_+ m_+}+ \delta_{n_- m_-}-2 \delta_{n_+ m_+}\delta_{n_- m_-}.
\end{align} 
Next we introduce an auxiliary
field $\phi\equiv \{\phi_{jn}\}=\{\phi_{jn_+n_-}\}$ carrying the same index structure as our $B$-fields,
and the quadratic action
\begin{align*}
    S[\phi]=-\frac{1}{4\Gamma} \phi_{jm} F_{mn}^{-1}M^{-1}_{jk}\phi_{kn},
\end{align*} 
defining the second moment
\begin{align}
    \label{eq:DiscretePhiContraction}
    \langle \phi_{jm}\phi_{kn} \rangle_\phi=-2\Gamma M_{jk} F_{mn}.
\end{align}
(As we need only this relation, there will be no need to compute the matrix inverses $M^{-1}$ and $F^{-1}$
explicitly.) A variable shift
\begin{align*}
    \phi_{jn}&\to \phi_{jn}+ 2\Gamma D \sum_{km} (M_{jk} F_{n,m}) (B_TB^\dagger)_{km}
\end{align*} 
generates a quartic term of the structure $(B_TB^\dagger)M\otimes F(B_TB^\dagger)$
cancelling the interaction vertex \eqref{eq:SintLowestOrder}, alongside the linear coupling
$\sum_{j,n}\phi_{jn} (B_TB^\dagger)_{jn}$. As a consequence, the full
quadratic action now reads
\begin{align*}
    S_0^{(2)}[B,\phi]=-D \sum_j \tr\left( B^\dagger_{n}dB_n  - B_n^\dagger  B_{Tn}\phi_{n}\right)_j,
\end{align*}
with the discrete time derivative $dB=B-B_T$, the `potential' $\phi$. This structure suggests
removing the field from the action by a gauge transformation of the $B$-field
variables, for which we try the ansatz 
\begin{align}
    \label{eq:BGaugeTransformationAbelian}
    B_{n}\to e^{\Theta_{n}}B_{n},\qquad B^\dagger_{n}\to B^\dagger_{n}e^{-\Theta_{n}}.
\end{align}
This transformation leaves the first term in the action invariant, and changes
the second as
\begin{align*}
   & B^\dagger_{n}B_{Tn}\to B^\dagger_{n}B_{Tn}e^{-d\Theta_{n}}\approx \\
    &\qquad\approx B^\dagger_{n}B_{Tn}(1-d\Theta_{n}),
\end{align*}
where $d\Theta_{n}= \Theta_{n}-\Theta_{n-1}$, or $d\Theta_{n_+ n_-}=\Theta_{n_+n_-}-\Theta_{(n_+-1)(n_-1)}$ in an
index resolved notation. Note the analogy to the
(symbolical) continuum formula $B^\dagger\partial_t B \to B^\dagger(\partial_t
-\partial_t \Theta)B$ representing a time-dependent gauge transformation. In the second equality we assumed smallness of the
interaction per time slice, $\textrm{var}(\phi)\sim \Gamma \sim 
\Lambda^2$,  
implying the smallness of $d\Theta_n$.  On the same basis,
$\phi\sim \Theta\ll 1$, allowing us to ignore the 
$\Phi$-dependence of the last term.
We conclude that the $\phi$-term is eliminated by solutions of the discrete time
difference equation $d\Theta=\phi$ (the analog of the continuum $\partial_t \Theta=\phi$) which we solve as 
\begin{align*}
    \Theta_n=\Theta_{n_+ n_-}=
            \sum\limits_{s=-\infty}^0 \phi_{n_++s,n_-+s},
\end{align*}
under the assumption that $\phi_{n_+,n_-}=0$, for negative discrete
time arguments. 
At this point, the interaction has been decoupled from the action, and we turn
to  the pre-exponential terms in the SFF \eqref{eq:FormFactorMultiSiteB}, which transform
as ($\Theta_{n_+n_-}$-notation from now on)
\begin{align}
       \label{eq:FormFactorInHS}
       K(t)&\to D^{2L} \sum_{\{m,n\}}\prod_j \left \langle (B_{tn} B_{n0}^\dagger B^\dagger_{0m} B_{mt})_j \right \rangle_B \times \nonumber\\
       &\qquad \left \langle e^{\sum_j(\Theta_{tm}-\Theta_{0m}-\Theta_{n0}+\Theta_{nt})_j} \right \rangle_\phi. 
\end{align}
Note the  shorthand notation,  $(B_{tn})_j\equiv B_{jtn_j}$ or
$(\Theta_{tn})_j=\Theta_{jtn_j}$, etc. Using that Eq.~\eqref{eq:BPropagatorDiscrete}
requires $t-n_j=m_j$, or $t-n=m$, in shorthand notation, the
exponent evaluates to
\begin{align}
    \label{eq:ThetaExponentDiscrete}
  &\Theta_{tn}-\Theta_{0n}-\Theta_{m0}+\Theta_{mt}=\\ 
  &\qquad \sum_{s=-t+1}^0 \phi_{(t+s)(n+s+t\Theta(s+n))}\equiv X[n,\phi].
\end{align}
To understand the structure of this expression, consider the  diagram shown in
Fig.~\ref{fig7} with the discrete time identifications
$t_+=t-n$ and $t_-=n$. Along the stretches $0\to t-n$ and $t-n\to t$   of duration $t-n$
and $n$, respectively, the time arguments of the  individual scattering amplitudes equals
$(t+s,t+n+s)$ and $(t+s,n+s)$.  The exponents sum the corresponding
time-dependent interaction fluctuation amplitudes, $\phi$.

In the final step, we need to integrate over $\phi$ in $ \langle e^{\sum_jX[n_j,\phi_j]}
\rangle_\phi$. This Gaussian integration yields
\begin{align*}
     & \langle e^{\sum_j X[n_j,\phi_j]}
    \rangle_\phi=e^{\frac{1}{2} \sum_{j,k} \langle X[n_j,\phi_j]X[n_k,\phi_k] \rangle_\phi}=\\ 
    & \quad =e^{-\Gamma\sum_{\langle j,k \rangle}  \sum_{s,u=-t+1}^0 
    F_{(t+s)(n_j+s+t\Theta(s+n_j)),(t+u)(n_k+u+t\Theta(u+n_k))}}\\
    &\quad=e^{-\Gamma t\sum_{\langle j,k \rangle}(1-\delta_{n_jn_k})}.
\end{align*}
In the last line, we noted that the function $F$ requires the pairwise equality
of at least two of its indices. For each of the  $s$-values in the double sum
the first two Kronecker-$\delta$ lock in for precisely one $u$-value, leading to
a contribution $\sum_s 2$. In that
remaining sum over $s$, we  need to take into account that for each pair of
nearest neighbors $\langle j,k \rangle$, and a given value of $t$, the brickwork
architecture implies $t/2$ two-qudit interaction processes, i.e. the summand is
non-vanishing for $t/2$ values of $s$ leading to the
damping rate  $\Gamma t$. Finally, for synchronous configurations $n_j=n_k$,
the product of two Kronecker $\delta$'s in the function $F$ balances the
damping, as indicated in the exponent.  
Summing over all configurations $\{n_j\}$, we arrive at the result \eqref{eq:FormFactorDiscreteTime}.

\section{Beyond the semiclassical approximation}
\label{app:BeyondSemiclassics}

In this Appendix, we provide  details concerning the analysis of the
discrete time path integral beyond leading order in perturbation theory. 

\subsection{Semiclassical exactness}
\label{app:SemiclassicalExactness}

Consider the form factor of a single qudit, Eq.~\eqref{eq:FormFactorSingle} as described by
an integral over $G=-iT \tau_3 T^{-1}$ with the action
Eq.~\eqref{eq:SigmaModelInvariant}. A series expansion of both, the functional
observable and the action in the generators $B$ defined through
Eq.~\eqref{eq:RationalParameterization} will yield the integral as a series in
powers of $D^{-1}$, but fall short of reaching perturbatively inaccessible
stationary points on the integration manifold. (Note the similarity of
Eq.~\eqref{eq:QMatrixB} with  the stereographic projection of
a sphere: an expansion around the north pole $Q=\tau_3$ or  $B=0$, will fail to reach the south
pole, $Q=-\tau_3$, or $B=\infty$.) `Semiclassical exactness' means that all
contributions to the expansion around individual stationary points beyond  Gaussian order
vanish. In the following, we review this feature for the expansion around the standard
saddle, however, the same principle works for all others. While discussions of
semiclassical exactness can be found in the literature (cf., e.g., Ref.~\cite{Haake}),
we here wish to demonstrate that the corresponding `higher order diagram
cancellations' survive the presence of interaction terms.

Demonstrating semiclassical exactness for the form factor as represented by
Eq.~\eqref{eq:FormFactorSingle} requires a separate expansion of the
pre-exponential observables and the action in generators, which is cumbersome.
To avoid this complication, we begin by switching to a generating function
representation.

Instead of working with `open boundary conditions' on the discrete time interval
$[1,t]$ as before, let us switch to one, still discrete, with periodic boundary
conditions, $t+1=1 \,\mathrm{mod} t$.  We also introduce  phase 
factors, $U\to  Ue^{i\phi_+}$, $U^\dagger\to  U^\dagger e^{-i\phi_-}$. With $\psi_{t+1}=\psi_1$ the path integral in
\eqref{eq:SingleQuditPartitionSum} then no longer yields unity but 
\begin{align*}
    Z(\theta)&  =\det(1-U^t e^{i \theta_+})\det(1-U^{\dagger t} e^{-i \theta_-}),
\end{align*}
with $\theta_\pm =t \phi_\pm $.
Expanding the determinants in these phase factors, we obtain the representation
\begin{align}
    \label{eq:KeldyshGeneratingFunction}
    K(t)=|\tr(U^t)|^2=\frac{1}{(2\pi)^2}\int_{0}^{2\pi}  d \theta_+ d \theta_- e^{-i (\theta_+-\theta_-) } Z(\theta).
\end{align}
Our task thus is to compute the generating functional in an expansion to leading
order in the factor $\exp(i (\theta_+-\theta_-))$.

Tracing the fate of the phases we find that they lead to a modification
$T_-B T_+ \to e^{i \phi} T_- B T_+ \equiv B_T$ in Eq.~\eqref{eq:S0DiscreteTime},
where $\phi=\phi_+ - \phi_-$. The inverse of the quadratic action defined
by the phase-decorated time translation operator is given by (cf. Eq.~\eqref{eq:BPropagatorDiscrete})
\begin{align}
    \label{eq:BPropagatorDamped}
    \Pi_{n,m}=\frac{1}{D}\delta_{\Delta n,\Delta m}
    e^{i \phi (\bar n-\bar m)}\left(\Theta(\bar n-\bar m)+\dots\right),
\end{align}
where the ellipses denote contributions with multiple windings around the
time circle, $\propto \exp(i k \theta)$, which do not contribute to the observable. The strategy now  is
to expand in nonlinear contributions to the action, compute terms in the ensuing
series by Wick's theorem and  extract the contribution $\propto e^{i \theta}$.

This task is
 simplified by the rule that contractions between neighboring
matrices $\tr(\dots B_{T } B^\dagger \dots  )$ vanish by causality: 
\begin{align*}
    \langle \tr(\dots B_{n_+-1, n_--1} B^\dagger_{n_- m_+} \dots  ) \rangle_B = \Pi_{n-1,n} (\dots)=0, 
\end{align*}
where we used that 
Eq.~\eqref{eq:BPropagatorDiscrete} enforces 
the incompatible conditions 
$n_+-m_+>2$ and $n_+=m_+$, 
and the vanishing of the
propagation one step backwards in time is prohibited by the temporal $\Theta$-function. (We here
ignore causality-allowed propagation around the periodic time cycle, $n\to
n-1$, as they are penalized by a damping factor $\exp(i \theta )$ and do not
contribute to the result at the required order in these factors.)

Inspection of action \eqref{eq:S0DiscreteTime} shows that some of our
contractions will involve the fields $dB$, acted upon by the discrete time
derivative (with phase factors included, i.e. $dB=B-e^{i \phi} T_-BT_+ $).) Since $d$ is the operator inverse of $\Pi$, cf.
Eq.~\eqref{eq:BPropagatorDiscrete} and the subsequent displayed equation,
$D(d\Pi)_{n,m}=\delta_{n,m}$, such Wick contractions yield the  simple
result
\begin{align}
    \label{eq:OBContractions}
     &\langle  \tr(A(dB) C B^\dagger ) \rangle_B= \frac{1}{D}\tr(A)\tr(C),\nonumber\\ 
    &\langle \tr(A(dB)) \tr(C B^\dagger ) \rangle_B=\frac{1}{D}\tr(A C).
\end{align} 
Physically, $d\Pi_{n,m}$ is vanishing, unless $n_+-m_+=n_--m_-=0$. These are
operator stretches of vanishing duration  (see Ref.~\cite{Muller2009} for their
in-depth discussion), which in an interacting context will
not be affected by interaction corrections. This observation establishes the
effectiveness of local semiclassical exactness, including in correlated multi
qudit systems. 

\begin{figure}[h]
    \centering
    \includegraphics[width=0.7\linewidth]{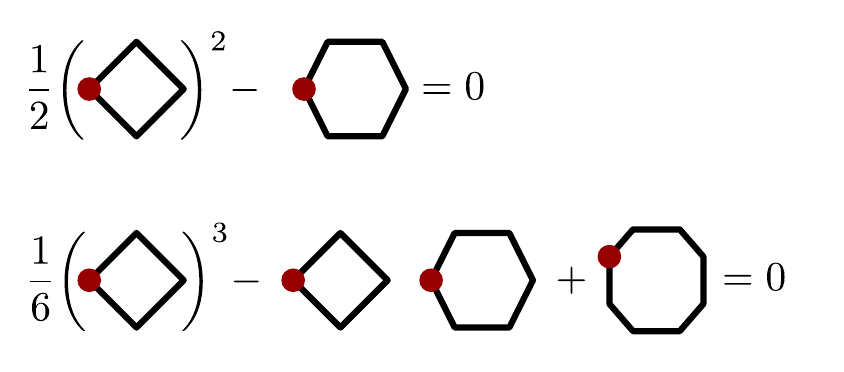}
    \caption{The cancellation of contributions to the $D^{-1}$-expansion of the
    functional illustrated for the two lowest orders, $D^{-2}$ (top) and
    $D^{-3}$ bottom. Polygons represent $\tr(B B^\dagger \dots B^\dagger dB)$
    traces with $2l$ corners matching the added  number of $B$- and
    $B^\dagger$-matrices. The vertex representing $dB$ is indicated by a red
    dot, and contractions are only possible   between vertices of
    different number parity relative to the dotted vertex, i.e.  between $B$'s
    and $B^\dagger$'s. The contraction between  the dotted $B$ in a polygon of degree $2l$ and
    any one of the $k$ $B^\dagger$-matrices in a $2k$-polygon yields $k$ identical $2(k+l-1)$ polygons, explaining the cancellation in the
    two lines of the figure.}
    \label{fig15}
\end{figure}

The contraction rules Eq.~\eqref{eq:OBContractions} are instrumental in
demonstrating semiclassical exactness. To understand how, consider the 
action Eq.~\eqref{eq:S0DiscreteTime} expanded
\begin{align*}
    S_0[B]&= D \sum_{l=0}^\infty (-)^l\,\tr((B^\dagger B)^l B^\dagger dB)\equiv \sum_l S^{(2l+2)}[B].
\end{align*} 
The expansion of these vertices out of the exponent leads to a series of
vertices which all  contain a single operator $dB$. To each
of these vertices, $S^{(2l+2)}$ we may associate a definite $D$-power in the
perturbative expansion. The structure of the Gaussian weight, $S^{(2)}[B]=D
\,\tr(B^\dagger d B)$ implies $B \sim D^{-1/2}$, and hence $S^{(2l+2)}\sim D
D^{-(2l+2)/2}=D^{-l}$

As an
example, consider the product
\begin{align*}
    &\left \langle (S^{(4)}[B])^2 \right \rangle=D^2\left \langle \left( \tr(B^\dagger B B^\dagger dB) \right)^2 \right \rangle\to \cr 
    &\quad\to 2D \,\tr((B^\dagger B)^2 B^\dagger dB)=-2S^{(6)}[B], 
\end{align*}
of two $D^{-1}$ operators, where the arrow means that we picked an operator $dB$
for contraction using the second of Eqs.~\eqref{eq:OBContractions}. This
one-step reduction exemplifies a general principle: the linkage of
vertices contributing to the expansion at a given order in $D^{-1}$ (presently
two $D^{-1}$-operators contributing at $D^{-2}$-level) leads to other operators
of the same order (a single $D^{-2}$ operator), showcasing
`perturbative renormalizability' of the model.

The instantaneous (in time) cancellation of all fluctuation diagrams in the
expansion of the functional can be demonstrated order by order in powers of
$D^{-1}$. Referring to Ref.~\cite{Muller2009} for the general proof,
Fig.~\ref{fig15} illustrates the cancellation for the orders
$D^{-2}$ (first line) and $D^{-3}$ second line. In this way, the expansion of 
the $B$-functional around
the standard saddle point, $B=0$, reduces to the quadratic functional with
action $S^{(2)}[B]$.

For completeness, we outline how this reduction leads to the semiclassical SFF. The final integration over the  reduced generating
functional yields $\det(d )^{-t}$, i.e. the $t$-th power of the determinant
of an operator
containing $1$  on the diagonal, and $\exp(i\phi)$ on the next-leading
diagonal. (The prefactor $D$ multiplying $d$ in the action cancels out due to the
normalization of the integral, $\int dB\exp(-D\, \tr(B^\dagger B))=1$.) A
straightforward calculation shows that  
$\det(d)^{-t}=\exp (-t\ln(1-e^{i \phi
t}))=\exp(t e^{i \theta}+\mathcal{O}(e^{2i\theta}))=1+ t e^{i \theta}+
\mathcal{O}(e^{2i \theta})$. 
Substituting this result into 
Eq.~\eqref{eq:KeldyshGeneratingFunction}, we obtain  $K(t)=t$. 

\subsection{Interaction vertex beyond leading order perturbation theory}
\label{app:InteractionBeyondQuartic}

We here demonstrate the vanishing of the interaction vertex on field
configurations satisfying the synchronization condition. To this end, consider
the expansion Eq.~\eqref{eq:InteractionVertexExpanded} of the interaction
between two neighboring qudits with Goldstone modes denoted by $B$ and $C$,
respectively. We have seen (cf. the discussion below Eq.~\eqref{eq:SintLowestOrder}) how the leading quartic
contribution to the interaction vanishes on the  Hadamard mode
$X_{mn}=(B\odot C)_{mn}$. 

In the remainder of this Appendix, we show that this
feature generalizes to all contributions to the sum above, and in this way
demonstrate that the Hadamard mode is a zero mode of the interaction
vertex. To this end, we decouple the quartic interaction via the
construction outlined in Section
\ref{sec:Interactions}. The removal of the interaction via a discrete time
dependent gauge transformation leads to the appearance of coherence-inducing
phase factors in all contributions to the action that cannot be expressed
solely in
terms of the $X$-mode. Focusing on late time physics, we ignore all these in the
following, thus effecting field reductions such as 
\begin{align*}
    &(B B^\dagger)_{nn} (CC^\dagger)_{nn}
    = B_{no}B^\dagger_{on} C_{np}C^\dagger_{pn}\to B_{no}B^\dagger_{on}C_{no}C^\dagger_{on}\\ 
    &\qquad =X_{no}X^\dagger_{on}=(XX^\dagger)_{nn}\\ 
    &(BB^\dagger)^p_{mn} (CC^\dagger)^q_{mn}\to \delta^{pq}(XX^\dagger)^p_{nm}.
\end{align*}
Using these rules, the  interaction vertex reduces to
\begin{widetext}
\begin{align*}
    S_{\textrm{i},1}[X]&= \frac{\Gamma D^2}{2}\sum_{n}\sum_{q=1}^\infty 
    {\rm tr}\left(
        (XX^\dagger)_{nn}^{q}+(X^\dagger
        X)_{nn}^{q} \right)=\Gamma D^2 \,\tr\left( \frac{XX^\dagger}{1-X X^\dagger}\right),\\ 
        S_{\textrm{i},2}[X]&= \frac{\Gamma D^2}{2}\sum_{n,m}\sum_{p,r=0}^\infty
         {\rm tr}\big((X X^\dagger)_{mn}^{p+1} (X
        X^\dagger)_{nm}^{r+1}+
        (X^\dagger X)_{mn}^{p+1}(X^\dagger X)_{nm}^{r+1}
        -\\
        &\qquad - (X (X^\dagger X )^p)_{mn} (X^\dagger ( XX^\dagger )^r)_{nm} - 
         (X^\dagger ( X X^\dagger )^p)_{mn} ( X ( X^\dagger X )^r)_{nm} \big)= \\ 
       &= \Gamma D^2\,\tr\left(XX^\dagger(X X^\dagger-1) \left(  \frac{1}{1-XX^\dagger} \right)^2   \right).
\end{align*}
\end{widetext}
It is easily checked that $S_{\textrm{i},1}[X]+S_{\textrm{i},2}[X]=0$, confirming the statement above.

\section{Variable change to the Hadamard mode}
\label{app:IntegrationHadamardMode}

To pass from an integration over $L$ synchronized $B$-field variables to that
over a single $X$ in Eq.~\eqref{eq:FormFactorBvsX},  we introduce a
Lagrange multiplier action
\begin{align*}
    S_\textrm{LM}[B,X,\Lambda]=\tr(\Lambda (X-\odot_j B_j))+ \tr((X^\dagger-\odot_j B^\dagger_j)\Lambda^\dagger ),
\end{align*} 
where the trace is over discrete time space. This locking leads to the
representation
\begin{align*}
    K(t)\to D^{2L} \int DX D \Lambda \, \left\langle e^{-S_\mathrm{LM}[B,X,\Lambda] }
     \right\rangle_B (X X^\dagger)_{t0} (X^\dagger X)_{0t},  
\end{align*} 
enabling the Gaussian integral  over the $B_j$'s with action
Eq.~\eqref{eq:S0DiscreteTime}. As a result we obtain  
\begin{align*}
   \left\langle e^{-S_\mathrm{LM}[B,X,\Lambda] } \right\rangle_B = e^{-\tr(X \Lambda + \Lambda^\dagger X)+
   \sum_{n,m}\Lambda_{n}\Pi_{nm}^L \Lambda^\dagger_m}, 
\end{align*} 
where the propagator is defined in Eq.~\eqref{eq:BPropagatorDiscrete}, and we
use the shorthand notation $n=(n_+,n_-)$. Crucially, the product of $L$ single
qudit  propagator matrix elements Eq.~\eqref{eq:BPropagatorDiscrete} equals $\Pi_{nm}^L
\equiv\Pi_{Lnm}$, where $\Pi_L$ is defined by Eq.~\eqref{eq:BPropagatorDiscrete} with
the replacement $D\to D^L$, i.e. the propagator of a single ergodic quantum
system with $D^L$ levels. This identification enables us to perform the
integration over $\Lambda$, leading to Eq.~\eqref{eq:FormFactorXQuadratic}.

\section{Including symmetry exchange correlations}
\label{app:SymmetryExchange}

We here describe how to advance from the symmetry exchange interaction
represented in quasiclassical Green functions Eq.~\eqref{eq:SSymmetriesG} to the
final result Eq.~\eqref{eq:SFFFormFactorSymmetries} (without relying on the
commutativity feature discussed in the main text.)
We start by using Eq.~\eqref{eq:GvsQDictionary}, where $Q=\{Q^{uv}\}$ and
$B=\{B^{uv}\}$ carry symmetry space structure, followed by leading order
expansion in generators to identify the quartic symmetry exchange vertex
\begin{align*}
  &S^{(4)}_\textrm{s}[B]\equiv -\frac{i \Gamma_\textrm{s}}{2} \sum_{\langle j,k \rangle}\quad\tr(B_TB^\dagger T^a)_{jnn}
  \tr(B_TB^\dagger T^a)_{knn}\\ 
    &\qquad - (B_T \leftrightarrow B^\dagger).
\end{align*}
We next decouple the  two traces contributing to the vertex  by a
Hubbard-Stratonovich field $\theta=\{\theta^{sa}_{jn}\}$, in analogy to our
discussion of the likewise quartic qudit correlation in
Appendix~\ref{app:HSInteraction}: 
\begin{align}
    \label{eq:ThetaAction}
        &S_{\textrm{s}}^{(4)}[B]\to \frac{1}{2i\Gamma_\textrm{s}} \sum_n   \Big(\sum_s s \theta^{saT}_{n}M \theta^{sa}_{n}+ \\
    &\qquad  D\theta^{+a}_{n}\tr(B_{T} B^\dagger T^a)_{nn}+ D \theta^{-a}_{n} (B^\dagger B_{T}T^{a})_{ nn}    \Big), \nonumber
\
\end{align}
where the matrix $M$ is defined in Eq.~\eqref{eq:MatrixMDef}. The second line
couples the symmetry generators into the quadratic $B$-action, which now reads
\begin{align*}
   & S^{(2)}_0[B,\theta]=D \sum_j  \times \\
    &\quad \tr\left( B^\dagger(B- B_{T}) +  B^\dagger (T^a \theta^{+a}) B_{T}+ B^\dagger B_{T}(T^a \theta^{-a})\right)_j.
\end{align*}
Remembering that $\theta=\{\theta_n\}$ carries structure in discrete time space,
this is again a case for elimination by discrete gauge transformation. We change
our fields in a manner resembling Eq.~\eqref{eq:BGaugeTransformationAbelian}, 
\begin{align*}
    B_{mn}\to U^+_m B_{mn}U^-_n,\qquad B^\dagger_{nm}\to \left( U^-_n  \right)^{-1}B^\dagger_{nm}
    \left( U^{+}_m \right)^{-1},
\end{align*}
where $U=\{U_j^{uv}\}$ are matrices in symmetry space, and carry a site index.
Substituting this ansatz into the action and requiring the elimination of
$\theta$ to first order in a small-$\theta$ expansion leads to the condition
\begin{align*}
   dU_n^+ = - U_n^+ (\theta^{+a}_n T^a), \qquad      dU_n^- = -  (\theta^{-a}_n T^a)U_n^-,    
\end{align*}
where we temporarily removed the site index for   transparency, and
$dU_n=U_n-U_{n-1}$. This is the discrete version of a non-abelian differential
equation, involving operators  not commuting at different
instances of time, $n$. 
It is solved by the likewise discrete version of an (anti-)time ordered exponential, 
\begin{align}
    \label{eq:TimeOrderedU}
    U^s_n = \textrm{T}_s \exp\left(- \sum_{m=0}^n \theta_m^{sa}T^a  \right),
\end{align}
where the time ordering protocol is defined as
\begin{align*}
    &\textrm{T}_+\exp\left( \sum_{m=0}^n X_n \right)\equiv \sum_k  \\ 
    &\qquad  \sum_{m_1=0}^n \sum_{m_2=0}^{m_1}\dots\sum_{m_k=0}^{m_{k-1}}
     \begin{cases}
        X_{m_k}X_{m_{k-1}}\dots X_{m_1},&s=+1,\\
        X_{m_1}X_{m_2}\dots X_{m_k},&s=-1.
    \end{cases}
\end{align*}
We are now in the same situation as before, the action has become
$\theta$-independent, while the source terms pick up a gauge factor.
Specifically, the SFF transforms as 
\begin{align*}
   K(t)&\to D^{2L} \prod_j \left \langle \tr(U^+_t B B^\dagger)_{j,t,0} \tr(U^-_t B^\dagger B)_{j,0,t} \right \rangle_{B,\theta}.
\end{align*}
Assuming that we are considering time scales $t\gg \Gamma^{-1}$, where
relaxation to the globally synchronized mode $X=\bigodot_j B_j$ has taken place, the integral over $B$ yields 
\begin{align*}
    K(t)=t \left \langle \prod_j\tr(U_{j,t}^+)\tr(U_{j,t}^-) \right \rangle_\theta. 
\end{align*}
Here, the product is over $2L$ independent traces
over  $\Bbb{C}^K$, where the matrices
$U^\pm_{j,t}=U^\pm_{j,t}(T^a)$ depend on the generators represented in $\Bbb{C}^k$. However, for our
purposes it is more natural to turn to a tensor space representation
\begin{align*}
    \prod_j \tr(U_{j,t}^\pm(T^a)) =\tr\left(\prod_j U_{j,t}^\pm(T^a_j)\right),
\end{align*}
where   the trace on the right-hand side now is
over the tensor space $(\Bbb{C}^K)^{\otimes L} $, and $T_j^a$, cf.
Eq.~\eqref{eq:TGeneratorTensorProduct}, are mutually commuting operators acting
in this space. With the evolution operators given by 
Eq.~\eqref{eq:TimeOrderedU}, we obtain
\begin{align*}
    K(t)=t \,\tr \left \langle \textrm{T}\,e^{-\sum_j\sum_{m=0}^t \theta_{jm}^{+a} T_j^a}  \right \rangle_{\theta^+}\times (+\leftrightarrow -),
\end{align*}
where the time ordering protocol applies to each of the $L$ exponentials 
individually. Expanding the time ordered exponential functions, and doing the
contractions of the $\theta$-variables according to the second moment 
\begin{align}
    \label{eq:ThetaSecondMoment}
    \langle \theta_{jm}^{r a} \theta_{kn}^{sb} \rangle= is \Gamma_\textrm{s} \delta^{rs}\delta^{ab}\delta_{mn}M_{jk},
\end{align} 
implied by Eq.~\eqref{eq:ThetaAction},  this yields Eq.~\eqref{eq:SFFFormFactorSymmetries}.

\section{Continuous time theory}
\label{app:ContinuousTime}

In this Appendix we detail the construction of various elements of the continuous time
theory. 

\subsection{Non-interacting theory}
To obtain the non-interacting part of the continuous time action, we substitute the
stationary phase configurations $G=T\bar G T^{-1}$ and $\Sigma = \lambda^2G$ 
into the action  \eqref{eq:ContinuumGSigma}. Noting  that the $T$-rotations
do not couple to  the second and
third term  in the first line due  cyclic invariance, we focus on the first
term. A leading order gradient expansion then yields 
\begin{align*}
     S_0[T]\equiv& -D\sum_j \tr \ln \left( i\partial_t - \lambda^2 T \bar G T^{-1} \right)_j=\\
    & =-D\sum_j \tr \ln \left( i\partial_t - \lambda^2 \bar G  - i T^{-1}[\partial_t,T] \right)_j\approx\\ 
    &\approx -\frac{D}{\lambda} \sum_j \tr\left( \tau_3 T^{-1} [\partial_t,T]\right)_j,   
\end{align*}
where in the second line we used the cyclic invariance of $\tr \ln$ and in the
third expanded  in the symmetry breaking operator
$T^{-1}[\partial_t,T]$ using the stationary phase equation
\eqref{eq:SCBAEquation} with the solution  $\bar G_{tt}=-is \tau_3$,
Eq.~\eqref{eq:GMeanFieldContinuousTime}. Representing the trace through time
integrals, we arrive at Eq.~\eqref{eq:SNoninteracingContinuous}.

\subsection{Interaction vertex}
\label{sec:AppInteractionContinuous}

Next, consider the interaction 
\begin{align*}
    &S_\textrm{i}[G]=    \frac{\Lambda^2 D^2}{2}\sum_{\langle j,k \rangle}
    \tr\left( (G_j\odot G_k)\tau_3 (G_j\odot G_k)\tau_3 \right)=\\
&=\frac{\Lambda^2 D^2}{2}\sum_{\langle j,l \rangle}\sum_s \int_{tt'}\\
&\left( (G_j\odot G_k)_{t t'}^{s s} (G_j \odot G_k)^{ss}_{t't} - 
(G_j\odot G_k)_{t t'}^{s \bar s} (G_j \odot G_k)_{t' t}^{s \bar s} \right). 
\end{align*}
We  substitute the representation $G=T \bar G T^{-1}$ with the mean field
solution Eq.~\eqref{eq:GMeanFieldContinuousTime} and the $T$ matrices given by
Eq.~\eqref{eq:RationalParameterization} into this action. The resulting
expressions can be organized according to the number of $B$-matrices. To zeroth
and second order, the action vanishes as a consequence of causality. These terms
contain the saddle point Green function in combinations $ (\bar G^s_{tt'})^2 (\bar G^s_{t't})^2$, which
evaluate to zero because $\Theta(s(t-t'))\Theta(s(t'-t))=0$. Terms of first and third order
 vanish because of the off-diagonal Keldysh block structure of
the matrix $T$.  The expansion thus starts at
 fourth order, as in the discrete time model: 
\begin{widetext}
\begin{align*}
   & S^{(4)}_\textrm{i}[B]\approx2\Lambda^2 D^2\sum_{\langle j,k \rangle} \int_{tt'}\\
   &\Big( \left((B[B^\dagger \bar G^+-\bar G^- B^\dagger])_{j,tt'} 
   (B[B^\dagger \bar G^+-\bar G^- B^\dagger])_{l,tt'}\right) \bar G_{t't}^{+2}+
    \left((B^\dagger[B \bar G^--\bar G^+ B)_{j,tt'}  
   (B^\dagger[B \bar G^--\bar G^+ B])_{k,tt'}\right) \bar G_{t't}^{-2}-\\
   & (B \bar G^--\bar G^+ B)_{jtt'}(B \bar G^--\bar G^+ B)_{ktt'}(B^\dagger \bar G^+-\bar G^- B^\dagger)_{jt't}
   (B^\dagger \bar G^+-\bar G^- B^\dagger)_{kt't}\Big), 
\end{align*}
\end{widetext}
where the first and second line represent the contribution of the term with
identical and opposite indices, i.e. the out- and the in-term respectively. 
We simplify these expressions by assuming slow variation of the $B$-fields,
which allows us to approximate their products with the quickly decaying mean field
Green functions as, e.g., 
\begin{align*}
    (B\bar G^-)_{t_+,t_-}=\int dt B_{t_+,t}\bar G^-_{t,t_-}\approx -\frac{i}{\lambda}B_{t_+,t_-},
\end{align*}
i.e. each $G^s$ in a non-causality violating expression enters as a factor $i
s/\lambda$. Similarly,  products of two
 Green functions featuring in the out-term, $\bar G^{\pm 2}_{t't}\propto \exp(- 2\lambda |t-t'|)\to
1/2\lambda$ upon time integration. In this way, we obtain the simplified
structure Eq.~\eqref{eq:SIntQuarticContinuum}.

\subsection{Summing over interaction processes}
\label{app:InteractionContinuousTime}
In this Appendix we adapt our treatment of interactions in
Appendix \ref{app:HSInteraction} to the continuous time framework. Essentially,
this amounts to taking continuum limits of the formulas discussed there.
We display them here for the sake of reusability of the formalism. Our starting
point is  the continuum action \eqref{eq:SIntQuarticContinuum} represented in a
manner analogous to Eq.~\eqref{eq:SIntCompactDiscrete}
\begin{align*}
            S_\textrm{int}^{(4)}[B]= \frac{2\Gamma D^2}{\lambda}\sum_{\langle j,k \rangle}\int_{tu}(BB^\dagger)_{jt} 
        F_{tu} M_{jk} (BB^\dagger)_{ku},    
\end{align*}  
where $(B B^\dagger)_t\equiv B_{t}B^\dagger_t\equiv B_{ t_+
t_-}B^\dagger_{t_- t_+}$ and  
\begin{align*}
   F_{tu}=\lambda^{-1}  \delta_{t_+,u_+}+\lambda^{-1}\delta_{t_-,u_-}- 8\lambda^{-2} \delta_{t_+,u_+}\delta_{t_-,u_-}. 
\end{align*} 
Next introduce the Hubbard-Stratonovich field $\phi_{jt}$ with action
\begin{align*}
    S[\Phi]=-\frac{\lambda}{8\Gamma  }\sum_{\langle j,k \rangle}\int_{tu} 
    \phi_{jt}M^{-1}_{jk}F^{-1}_{tu}\phi_{ku},
\end{align*}
and resulting second moment 
\begin{align}
    \label{eq:PsiContinuumExpectation}
    \left \langle \phi_{jt} \phi_{k u} \right \rangle =-4\Gamma M_{jk} 
    F_{tu}.
\end{align}
A shift  $\phi_{jt} \to \phi_{jt}+ \frac{4\Gamma D}{\lambda}  \int_u \sum_{k} F_{tu}M_{jk} (BB^\dagger)_{ku}$ removes the
quartic interaction but adds a linear $\phi (BB^\dagger)$ coupling, so that the
quadratic action is now given by
\begin{align*}
    S^{(2)}[B,\Phi]=D\sum_j \int_{t}B^\dagger_{jt}\left(\frac{2}{\lambda}\partial_t - \phi_{jt} \right)B_{jt},
\end{align*}
where the time derivative acts on the center coordinate: $\partial_t B_t=
(\partial_{t_+}+ \partial_{t_-}) B_{t_+,t_-}$. We remove the `potential' $\phi$ via the gauge transformation
\begin{align*}
    B_{t}\to e^{\Theta_{t}}B_{t},\qquad B^\dagger_{t}\to B^\dagger_{t}e^{-\Theta_{t}},
\end{align*}
with 
\begin{align*}
    \Theta_t=\Theta_{t_+ t_-}=\frac{\lambda}{2}\int_{-\infty}^0 ds \, \phi_{t_++s,t_-+s}.
\end{align*} 
The continuum time analog of the gauged form factor \eqref{eq:FormFactorInHS}
then reads
\begin{align}
       \label{eq:FormFactorContinuousTime}
       K(t)&\to \frac{D^{2L}}{\lambda^{3L}} \int_\tau \prod_j \left \langle (B_{t \tau} B_{\tau 0}^\dagger B^\dagger_{0 (t-\tau)} B_{(t-\tau) t})_j \right \rangle_B \times \nonumber\\
       &\qquad \left \langle e^{\sum_j(\Theta_{t \tau}-\Theta_{0\tau}-\Theta_{(t-\tau) 0}+\Theta_{ (t-\tau) t})_j} \right \rangle_\phi, 
\end{align}
where $\tau=\{\tau_j\}$ defines the set of qudit time integration variables (cf.
Fig.~\ref{fig7}). We simplify the exponent in analogy to Eq.~\eqref{eq:ThetaExponentDiscrete}
 \begin{align*}
   &\Theta_{t \tau}-\Theta_{0\tau}-\Theta_{(t-\tau) 0}+\Theta_{ (t-\tau) t}=\\ 
   &\qquad=\frac{\lambda}{2}\int_{-t}^0 ds\,\phi_{t+s,\tau+s+t \Theta(s+\tau)} \equiv X[\tau,\phi],
 \end{align*} 
and hence obtain for the averaged exponent, 
\begin{align*}
         & \langle e^{\sum_j X[\tau_j,\phi_j]}
    \rangle_\phi=e^{\frac{1}{2} \sum_{j,k} \langle X[\tau_j,\phi_j]X[\tau_k,\phi_k] \rangle_\phi}=\\ 
    & \quad =e^{-\frac{\Gamma \lambda}{2}\sum_{\langle j,k \rangle}\int_{-t}^t ds du\, 
    F_{(t+s)(\tau_j+s+t\Theta(s+\tau_j)),(t+u)(\tau_k+u+t\Theta(u+\tau_k))}}\\
    &\quad=e^{-\Gamma  t\sum_{\langle j,k \rangle}\left(1-\frac{8}{\lambda}\delta_{\tau_j-\tau_k}\right)}. 
\end{align*}
We finally recall that all  $\delta$-functions  are
smeared over intervals $\sim \lambda^{-1}$. We also know that for a  synchronized
configuration $\theta_j=\textrm{const.}$, interactions will not dephase the form
factor. Use this condition to fix the normalization of the last remaining
$\delta$-function as $1=8\lambda^{-1}\delta_0$, we identify  the
exponent with the function specified below Eq.~\eqref{eq:SFFTimeFunctional}. 

In a final step, we compute the average
$\langle \dots \rangle_B$ in Eq.~\eqref{eq:FormFactorContinuousTime}. 
Each of the contractions of fields $B_j$
 over the action
Eq.~\eqref{eq:S0Continuum} brings down a factor $\sim\lambda^2/D$. Absorbing a factor $\lambda^L$ in
the measure of integration over the  time arguments $\tau_j$, we
arrive at Eq.~\eqref{eq:SFFTimeFunctional}.


\end{document}